\documentclass[11pt]{article}
\usepackage[a4paper]{geometry}
\usepackage{graphicx}
\usepackage[english]{babel}
\usepackage[round]{natbib}
\usepackage{todonotes}
\usepackage{amsmath}
\usepackage{hyperref}
\usepackage{tabularx}
\usepackage[table]{xcolor}
\usepackage{booktabs}
\usepackage{threeparttable}
\usepackage{soul}
\usepackage{float}
\usepackage{subcaption}
\usepackage[utf8]{inputenc}
\usepackage{authblk}
\usepackage{booktabs}
\newcommand{\coo}{\ensuremath{\mathrm{CO_2}}}

\begin{document}

\begin{center}

\vspace*{0.8cm}

{\LARGE\bfseries
Economic Benefits of Cross-Border Electricity Trade and\\[0.2em]
Domestic Electricity Production in Switzerland
\par}

\vspace{1.0cm}

{\large
Benjamin Lietti$^{1,*},$ \quad
Ambra Van Liedekerke$^{2,*},$ \quad
Marino Bucher$^{3,*},$ \quad
Michael Beer$^{1},$ \quad
Nicolas Eschenbaum$^{4},$ \quad
Reto Foellmi$^{3},$ \quad
Blazhe Gjorgiev$^{2},$ \quad
Fabian Gottschlich$^{1},$ \quad
Giovanni Sansavini$^{2},$ \quad
Christian Schütz$^{1},$ \quad
Urs Trinkner$^{4}$
\par}

\vspace{0.5cm}

{\small
$^{1}$ BKW Energie AG \qquad
$^{2}$ ETH Zurich \\
$^{3}$ University of St.~Gallen \qquad
$^{4}$ Swiss Economics
\par}

\vspace{0.3cm}

{\footnotesize
$^{*}$ Corresponding authors:
\href{mailto:benjamin.lietti@bkw.ch}{benjamin.lietti@bkw.ch},
\href{mailto:avanliede@ethz.ch}{avanliede@ethz.ch},
\href{mailto:marino.bucher@unisg.ch}{marino.bucher@unisg.ch}
\par}

\vspace{0.5cm}

{\small \today}

\end{center}

\vspace{0.8cm}

\noindent\rule{\textwidth}{0.4pt}

\vspace{0.4cm}

\begin{abstract}
This paper assesses the economic benefits of cross-border electricity trade in the context of rising energy prices and geopolitical uncertainty. Switzerland provides a particularly compelling case because of its geographical location and limited integration into the EU electricity market. We first use the Nexus-e framework to simulate the generation capacity and transmission grid expansion required under autarky and free-trade scenarios. Based on these capacity configurations, a fundamental power market model simulates wholesale market outcomes under projected demand and generation conditions for 2040. By jointly accounting for consumer and producer surplus, as well as investments in generation capacity and grid infrastructure, we provide a comprehensive assessment of the welfare effects of cross-border electricity trade, leading to an estimated EUR 1.5 billion gain. More specifically, open access to cross-border electricity trade implies a EUR 1.8 billion transfer from largely publicly owned producers to consumers.  Finally, we provide an illustrative, non-causal mapping of broader macroeconomic effects. Under a price-convergence assumption, the implied demand adjustment corresponds to a GDP-equivalent effect of approximately EUR 7 billion.
\end{abstract}

\vspace{0.3cm}


\vspace{0.2cm}


\section{Introduction}\label{sec:intro}
 
Cross-border electricity trade is of critical importance to Switzerland. Over the 2015–2025 period, the country had a positive power trade balance, with net exports averaging about 1.7 TWh per year.\footnote{ Data provided by Bundesamt für Energie BFE.} Cross-border power trading generates significant economic value for Switzerland and for its flexible hydropower capacity in particular, allowing for the optimal utilization of power infrastructure. In addition, international trade strongly supports security of supply. However, considering only the average yearly trade balance provides only a partial picture of the Swiss power system and can be misleading regarding the country’s energy autonomy. When restricting the trade balance to the winter months, the data reveal a different situation. With net winter imports averaging 4.2 TWh, the Confederation relies heavily on the European power market to satisfy domestic electricity demand. At the end of winter, remaining hydropower reserves averaged 1.5 TWh, representing only one third of the required imports and highlighting the crucial importance of access to international electricity markets. 

\begin{figure}[h]
    \centering
    \includegraphics[width=1\linewidth]{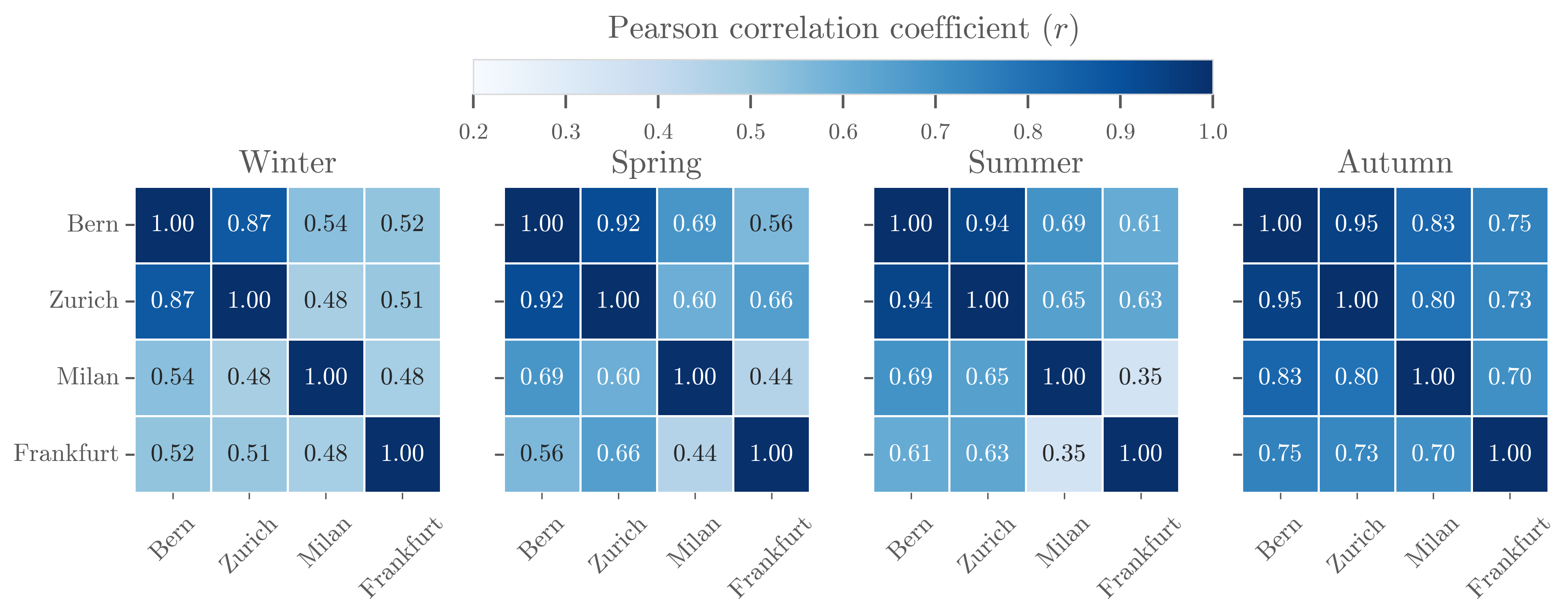}
    \caption{Solar irradiance correlation over the year 2025.}
    \label{fig:sun_correlation}
    \footnotesize
    \textit{Notes:} The figure shows the regional correlation of total daily solar irradiance over the year 2025. A higher value corresponds to more similar PV production profiles.

\end{figure}

Power trade also strongly supports renewables deployment necessary to close the Swiss winter power gap in a sustainable manner. Wind and solar generation depend on weather conditions and are not fully dispatchable. Consequently, meteorological fluctuations or prolonged adverse conditions can significantly affect grid stability and electricity prices. One way to mitigate this risk is to improve access to geographically diversified generation resources and cross-border interconnection. Figure  \ref{fig:sun_correlation} supports this argument by outlining the correlation of total daily solar irradiation over the year 2025 across several Swiss and European cities. In addition, the seasonal decomposition illustrates that the greatest diversification benefits would occur during the winter months, coinciding with strained grid periods in Switzerland.

The other viable solution to close the winter gap is to rely more heavily on flexible generation such as gas or oil power plants. This alternative strategy implies massive additional investments, as the existing capacity would be insufficient for most countries. In addition, pursuing this strategy would ultimately lead, without counter-measures such as carbon dioxide removals for instance, to a significant increase in greenhouse gas emissions and higher electricity prices. Given the long lead times required to develop new hydropower capacity, from initial conception to final implementation, it does not represent a feasible solution to address this issue within the timeframe of our analysis.

These strategic trade-offs are well captured by the “impossible energy trinity” framework proposed by \cite{thaler_impossible_2022}, which describes a trilemma between sustainability, security of supply, and sovereignty. These three objectives cannot be achieved simultaneously, and security of supply is generally treated as a binding policy objective for an advanced economy. The situation therefore presents a dilemma: either sacrifice some sovereignty to enhance power-system security or accept a less sustainable pathway.

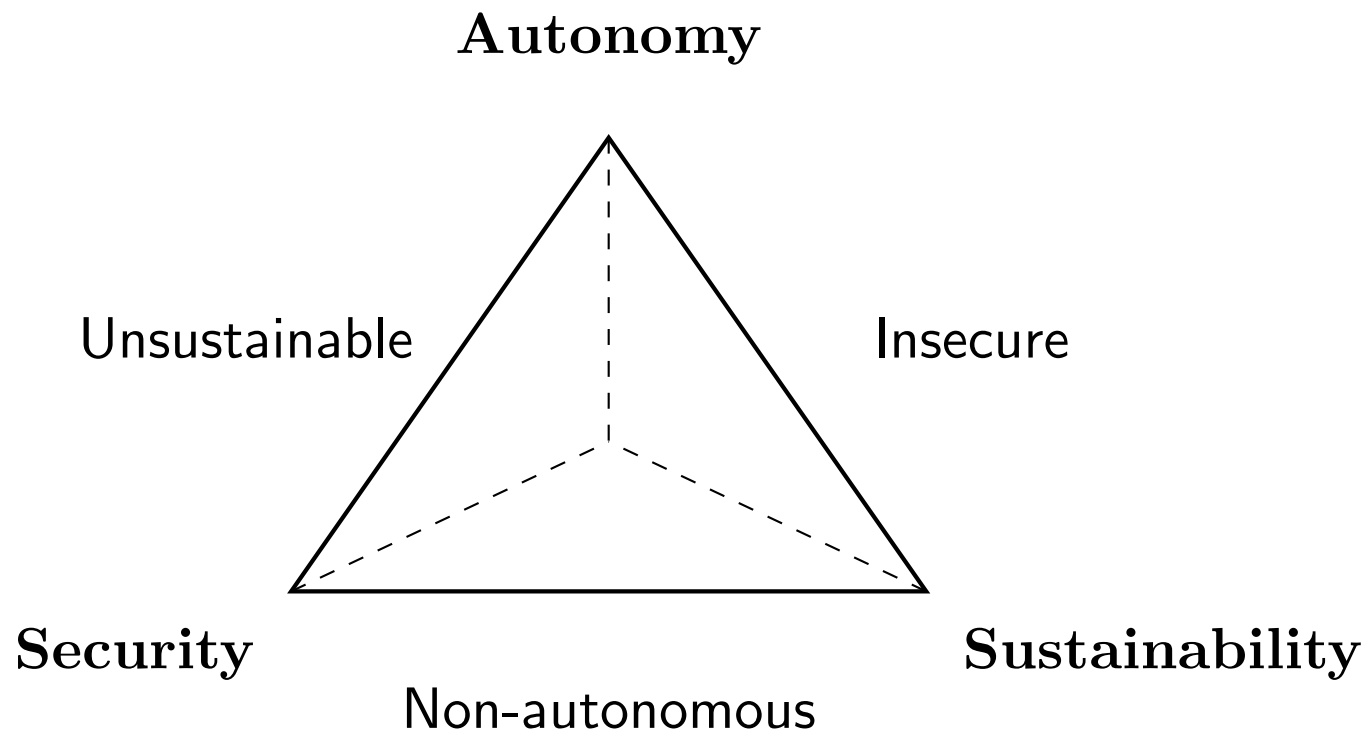
\begin{figure}[ht]
\centering
\begin{tikzpicture}[
    scale=0.6,
    every node/.style={font=\sffamily}
]

\coordinate (A) at (0,5);
\coordinate (B) at (-3.5,0);
\coordinate (C) at (3.5,0);

\coordinate (O) at (0,1.65);

\draw[thick] (A) -- (B) -- (C) -- cycle;

\draw[dashed] (A) -- (O);
\draw[dashed] (B) -- (O);
\draw[dashed] (C) -- (O);

\node[above=0.35cm, font=\bfseries]
    at (A) {Autonomy};

\node[below left=0.15cm, font=\bfseries]
    at (B) {Security};

\node[below right=0.15cm, font=\bfseries]
    at (C) {Sustainability};

\node at (-4,2.8) {Unsustainable};
\node at (4,2.8) {Insecure};
\node[below=0.5cm] at (0,0)
    {Non-autonomous};

\end{tikzpicture}
\caption{Trade-offs between Autonomy, Security and Sustainability \citep{thaler_impossible_2022}}
\label{fig:energy_triangle}
\end{figure}

Against this background, this paper evaluates the economic and financial implications of pursuing a policy of full autarky, compared to an open-trade scenario under comparable security-of-supply constraints. To ensure transparent and reproducible results, we rely on simulation models, and, more specifically, the outcomes of a fundamental power market model as the cornerstone of our analysis. Switzerland is a particularly suitable case study for several reasons. First, the country, while being geographically located in the heart of Europe, is not integrated into the European power market or the more efficient Flow-Based Market Coupling system and instead relies on bilateral and multilateral agreements. In other words, Switzerland currently lies in an intermediate and temporary situation. Second, the power landscape of Switzerland, with significant hydropower storage capacity and its lack of sufficient domestic production during the winter months, makes it a particularly compelling case for our research question. Third, the relatively small size of its electricity market compared with those of neighboring countries allows us to abstract from second-order international effects. Finally, reliable benchmarks based on the existing literature are available. For instance, the
Association of Swiss Electricity Companies \citep{vse_energiezukunft_2025} argues that an electricity agreement
with the EU would enhance supply security and reduce costs, while the
lack thereof limits access to cross-border capacities and increases prices. 

To this end and assuming no deviation from the current electricity generation and demand development in neighboring European markets, we construct two scenarios based on Net Transfer Capacities (\textit{NTC} hereafter). The Trade scenario features expanded cross-border transfer capacity, whereas the Autarky scenario assumes complete isolation. Within these environments, we use the CentIv and Cascades models from the Nexus-e framework \citep{Gjorgiev_Nexuse_2022} to simulate the cost-efficient development of generation capacity required to meet demand and the grid expansion within Swiss borders that is needed to achieve satisfactory power system security, respectively. Such simulations are made for the year 2040, which we use as the reference year.\footnote{This reference year has been chosen based on the Swiss nuclear phase-out program, assuming a 60-year lifetime. Also, a 14-year horizon makes the adaptation investments in grid and generation realistic, while falling within a reasonably foreseeable planning horizon, allowing for more reliable assumptions about future electricity-market conditions.} The resulting generation and transmission configurations are then used as inputs to BKW’s fundamental price-forecasting model to simulate wholesale market outcomes. This process yields hourly power demand and prices for both the autarky and trade scenarios in 2040, providing the basis of our economic analysis.

First, to ensure transparency and traceability, we focus on direct and observable effects for agents on the electricity market. Accordingly, this first-step analysis adopts a short-run perspective in which electricity demand is held fixed across the Trade and Autarky scenarios, abstracting from longer-run demand and economic adjustments; this assumption is subsequently relaxed in the second part of the analysis, which considers the potential macroeconomic response to persistent electricity-price differences. Although potentially substantial, externalities such as locally deteriorated air quality associated with expanded gas-fired capacity, as well as broader general-equilibrium effects are therefore not considered at this stage. Hidden costs of joining the EU power market for Swiss companies may also arise from legal barriers or institutional asymmetry between EU members and Switzerland such as the entry of foreign competitors on the domestic market \citep{heselhaus_quest_2022}. Neglecting these aspects, our results indicate that the autarky scenario entails a social welfare loss of EUR 1.5 billion in 2040, relative to the trade scenario. Only around 8\% of this cost is attributable to the additional generation and transmission investments required. Consumers bear the largest direct burden, with consumer surplus declining by an estimated EUR 3.2 billion, driven by a higher average market price. However, these losses are partly offset by gains for domestic producers, who benefit from limited imports and higher prices, yielding an estimated increase in surplus of EUR 1.8 billion.

In a second step, we complement the electricity-market welfare analysis with an illustrative assessment of broader macroeconomic consequences. Building on a price convergence assumption, we assume that the long-run load-weighted average electricity price under autarky converges to the price under trade. We therefore reduce electricity demand under autarky until the two average prices match. We then map the implied reduction in electricity consumption into economic activity using the historical relationship between temperature-adjusted electricity consumption and GDP. Under this framework and using demand variation as a proxy for GDP, results suggest an estimated GDP-equivalent effect of approximately EUR 7 billion. Importantly, this GDP-equivalent effect captures a different economic object from the electricity-market welfare effect estimated in the first step and should therefore not be interpreted as an additional welfare loss or added to the latter.

The remainder of this paper is organized as follows: Section \ref{sec:literature} covers the existing literature; Section \ref{sec:method} details the methodology applied and the different models used. Section \ref{sec:study} develops the methods used to estimate the economic consequences; Section \ref{sec:results} presents the results and discusses the economic costs of a total autarky for Switzerland. Finally, Section \ref{sec:conclusion} concludes with a summary of the main findings and open discussion.

\section{Literature Review}\label{sec:literature}

This paper relates, on the one hand, to the extensive literature investigating the tight links between energy and economics. \cite{bohringer_energy_2018} discuss the economy-wide rebound effect of increases in energy efficiency, emphasizing the significant impact of energy on economic activity. In addition, the paper provides a general description of the different channels through which energy shocks propagate. More closely related to our study, \cite{iea_building_2025} discusses the necessity of building an electricity transmission grid that allows systems to deal with the increasing demand for electricity. The report shows that an enhancement of grid infrastructure is required to facilitate cross-border interconnections and the deployment of wind and solar power. The large expected increase in electricity consumption, driven by industrial development, electrification of heating and transport, as well as new technologies such as artificial intelligence, creates the need for comprehensive analyses of the electricity network. Against this background, \cite{allard_european_2020} evaluate the transmission grid reinforcement required to integrate variable renewable energy sources on a large scale using a transmission grid module to model the energy system. \citet{iea_world_2022} estimates that transmission grid investment requirements are projected to be approximately three times higher than distribution grid investments in advanced economies over the period 2031–2040, whereas a ratio of roughly two is expected for developing and emerging economies. \citet{raycheva_risk-informed_2023} model the Swiss transmission system and conclude that a mere cost-minimization objective in planning the grid infrastructure in Switzerland does not provide sufficient system security. The main contribution of the present study lies in its comprehensive accounting of the infrastructure required to ensure security, as well as the effects of the power plant stack composition on electricity prices and generation costs. 

This work also relates to the literature on cross-border transmission capacities. In this context, we aim to contribute to the discussion of the risks and benefits of electricity trading. Full integration in the EU power market can potentially cause increased administrative costs and difficulties. In fact, the institutional framework has evolved through successive EU energy directives, and so has the regulatory path toward market integration, as outlined by \citet{pollitt_european_2019}. Legal barriers are further discussed by \citet{heselhaus_quest_2022}, who highlights the institutional asymmetry between EU members and Switzerland in decisions regarding energy mix and governance. These associated costs must be weighed against several well-documented benefits of electricity trade. For example, \citet{hofmann_role_2022} point out that Switzerland’s exclusion from Flow-Based Market Coupling leads to grid congestion and unscheduled flows, reducing the country’s influence over European  electricity governance despite its importance as a transit hub; \citet{newbery_benefits_2016} simulate electricity prices before and after market coupling in Europe, showing that integration improves market efficiency primarily through better balancing services and intraday coordination; \citet{von_der_fehr_water_1997} demonstrate that trade between hydro-based and thermal-based systems allows for cost savings through intertemporal substitution of electricity. Such results demonstrate that cross-border electricity trade offers significant efficiency gains by enabling the optimal allocation of generation resources across countries, with complementarities across generation technologies further amplifying these benefits.

 Nevertheless, while previous papers investigate market rigidities and electricity trade, the present study goes further, exploring the impact on prices at a more disaggregated level using scenario analyses. In this sense, it places us closer to \citet{acer_acers_2022}, which estimates that integration has brought annual benefits of around EUR 34 billion across Europe, based on comparisons to a no-trade baseline. Finally, broader economic impacts of grid infrastructure are emphasized by the International Energy Agency \citep{iea_electricity_2023}, which stresses that insufficient investment in transmission networks hampers the full realization of trade benefits and increases system costs, with current investment falling short of what is needed for a sustainable energy transition.

Closer to our work and in a more detailed bottom-up modeling effort, \citet{eser_trade-offs_2018} simulate Switzerland’s electricity future under two scenarios and find that stronger integration with Europe reduces costs but increases regulatory and investment risks. While the authors investigate the impacts of increased interconnections against a \textit{status quo} situation, our study goes further, simulating the consequences of complete exclusion from the EU power market.

We first quantify welfare effects using simulated market outcomes and then extend the framework by developing a more conceptual framework, linking electricity consumption with GDP to eventually estimate the macroeconomic impact of autarky. This approach relates our paper to the literature examining the close relationship between electricity trade and consumption, and economic activity. \citet{antweiler_cross-border_2016} presents a model with asymmetric costs and transmission frictions, showing that trade volumes increase with demand and capacity utilization, but decline with distance and limited transmission capacity. An empirical analysis by \citet{batalla-bejerano_energy_2019} uses a gravity equation framework to confirm that trade flows follow patterns consistent with economic size, distance, and institutional arrangements such as free trade agreements. 

The use of electricity consumption as a proxy for economic activity has a long tradition in the literature dating back to the 1970s–1990s. However, the increased access to timely and detailed official economic data reduced interest in such information until the late 2010s when new machine learning models and highly granular, real-time data expanded analytical possibilities, ultimately proving decisive during the COVID-19 pandemic. Our paper contributes to this renewed strand of literature by exploiting electricity consumption to estimate the macroeconomic impact of autarky. In this vein, we build on the findings of \citet{aroraindocator}, who show that, despite structural and technological change, electricity consumption and GDP remain tightly related. \citet{JanzenCovid} use the decrease in electricity consumption in Switzerland during the COVID-19 crisis as a real-time indicator of the economic shock. More specifically, we follow \citet{fezzi2020real} and correct for the impact of temperature on load demand before estimating economic impacts.

Overall, despite the significant uncertainty surrounding the costs and benefits of cross-border trading for Switzerland, recent studies have examined the economic relevance of Swiss participation in the European electricity market. \citet{frontier2021stromzusammenarbeit} evaluates several forms of CH–EU electricity cooperation for 2025 using a coupled power market and network modeling framework. The study finds that reduced cooperation weakens Swiss security of supply and lowers welfare. Our paper complements such results by focusing on long-term outcomes. This allows us to consider impacts of a substantially different environment, characterized by increased electrification, new consumption segments such as data centers as well as changes in generation technologies, with greater renewable generation capacity and lower nuclear capacity.  More recently, \citet{ecoplan2025stromabkommen} assesses the macroeconomic implications of the negotiated electricity agreement. Its analysis combines a static electricity-trading model with a multi-country general-equilibrium model, finding that secured market coupling can generate additional trading gains, lower system costs, reduce electricity prices, and increase GDP.

Our paper complements the literature in three ways: \textit{i)} we endogenously model generation and transmission expansion under fixed security conditions and use a fundamental hourly power-market model to quantify distributional effects across consumers and producers; \textit{ii)} we quantify the impact of a worst-case scenario, characterized by a fully isolated electricity system; \textit{iii)} we develop an electricity-demand-based framework to derive an evaluation of broader GDP-equivalent effects, thereby linking market-level price effects to potential economy-wide consequences.

\section{Methodology}\label{sec:method}

\subsection{Nexus-e models} 
The study aims to compare the outcomes of two scenarios for the reference year 2040: one characterized by complete integration into the European power market (Trade) and the other representing total autarky (Autarky). 
To do so, we model the future electricity system under these scenarios using Nexus-e, a framework of soft-linked models that simulate different aspects of the energy system~\citep{Gjorgiev_Nexuse_2022}.
In particular, the Nexus-e models used for this study are CentIv and Cascades.

CentIv is an optimal generation expansion and dispatch model. 
As such, it computes cost-optimal investments in generation technologies, as well as their operation, to satisfy the electricity demand, while respecting generation, storage, and network constraints~\citep{Raycheva_CentIv_2020}. 
Alongside generation capacity, transmission capacity is a cornerstone of the power system. 
As shown by \citet{abrell_cross-country_2016} using a general equilibrium model, transmission expansion, rather than renewables alone, is the key driver of increased trade and efficiency in Europe. 
For that reason, we make use of Cascades, a cascading failure model, to compute system security, taking as inputs the optimal generation mix and dispatch computed by CentIv.
Additionally, Cascades identifies the optimal transmission expansion plan that achieves security levels comparable to the 2018 reference case reported by ~\cite{raycheva_risk-informed_2023}.
Each line and transformer already present in the transmission system can be doubled with a new identical line or transformer.
\footnote{Transmission expansion in Cascades is restricted to existing corridors through parallel duplication of identical lines. Greenfield connection candidates are excluded. This assumption maintains computational tractability while mirroring standard industry practices, where expanding existing rights-of-way is prioritized over new corridor acquisition.}
Grid elements are selected for expansion based on their impact on demand not served (DNS) and line overloads, in an iterative procedure, as explained in~\cite{raycheva_risk-informed_2023}.
The iterative procedure terminates when the reference risk is achieved.

The Nexus-e CentIv and Cascades models require input data detailing the Swiss power system's generation and storage units, electricity demand, and the transmission network. 
The input data for the Nexus-e framework are described in~\cite{Nexuse_input_v1}.
However, they have been updated with more recent information on the generators' investment and operating costs, the currently installed capacities, the planned capacity expansions in neighboring countries, the electricity demand, and the network topology.

Generation investment and operating costs for 2040 are updated using more recent estimates from the literature.
Tables~\ref{tab:inv_cost}, \ref{tab:FOM_cost}, and \ref{tab:VOM_cost} report information on the applied investment costs, fixed operating and maintenance (FOM) costs, and variable operating and maintenance (VOM) costs, respectively.
The current power system's generation mix is calibrated according to the information in~\cite{BFE_elekrizitätsproduktionsanlagen} and~\cite{entsoe_power_stats} for Switzerland and its neighboring countries, respectively.
The generation expansion of the neighboring countries is defined by~\cite{TYNDP2024}.
Swiss generation expansion is instead optimized by the CentIv model, ensuring the annual generation target from renewables of 39.2 TWh is satisfied in 2040, according to the Federal Act on a Secure Electricity Supply from Renewable Energy Sources~\citep{bfe_federal_act}. While politically grounded and defensible, this constraint may artificially limit gas-fired generation expansion under autarky, thereby dampening the resulting increase in electricity prices and yielding a more conservative estimate of the costs of autarky.
All hydropower plants are modeled with an availability factor computed as the average availability factor for the years 2020--2024, according to~\cite{eex_availability_factor}.
The 2040 annual demand is set to that in~\cite{energieperspektieven2050} and increased by projections of data center electricity demand based on different reports.
The current Swiss transmission grid is modeled according to data from~\cite{swissgrid_topology}.
Additionally, the 2040 grid includes the grid expansions planned by 2025, as detailed in~\cite{swissgrid_strategic_grid}.
Grid expansions come at a fixed cost per km and per MW for line upgrades and transformer upgrades, respectively, as detailed in~\cite{raycheva_risk-informed_2023}.
The connection to the neighboring countries is modeled using the existing cross-border transmission lines.
However, the cross-border power exchange is limited by the NTC values detailed in Table~\ref{tab:ntc_ch}. 
The Autarky scenario is modeled by setting NTC values between Switzerland and its neighboring countries to 0.
Furthermore, the parameters of the cross-border lines are modified to avoid power flowing through them.

\begin{table}[h]
    \begin{center} 
    \caption{Net Transfer Capacities (NTC) for Switzerland's cross-border connections in 2040.} 
    \label{tab:ntc_ch}
    \begin{tabular}{llcccc} \toprule
        Scenario & CH-DE & CH-FR & CH-AT & CH-IT \\ \midrule
        Trade & 7100 & 4000 & 1200 & 4100 \\
        Autarky & 0 & 0 & 0 & 0 \\
        \bottomrule
    \end{tabular}
    
    \footnotesize
    \textit{Notes:} All values are in MW. The 2040 values are defined as the maximum observed import values in 2024, expanded to account for the TYNDP 2024 projects \citep{TYNDP2024}.
    \end{center}
\end{table}

\subsection{Generation and transmission expansion}
\begin{figure}[H]
    \centering
    \includegraphics[
        width=1\linewidth,
        trim=0 0 0 1.15cm,
        clip
    ]{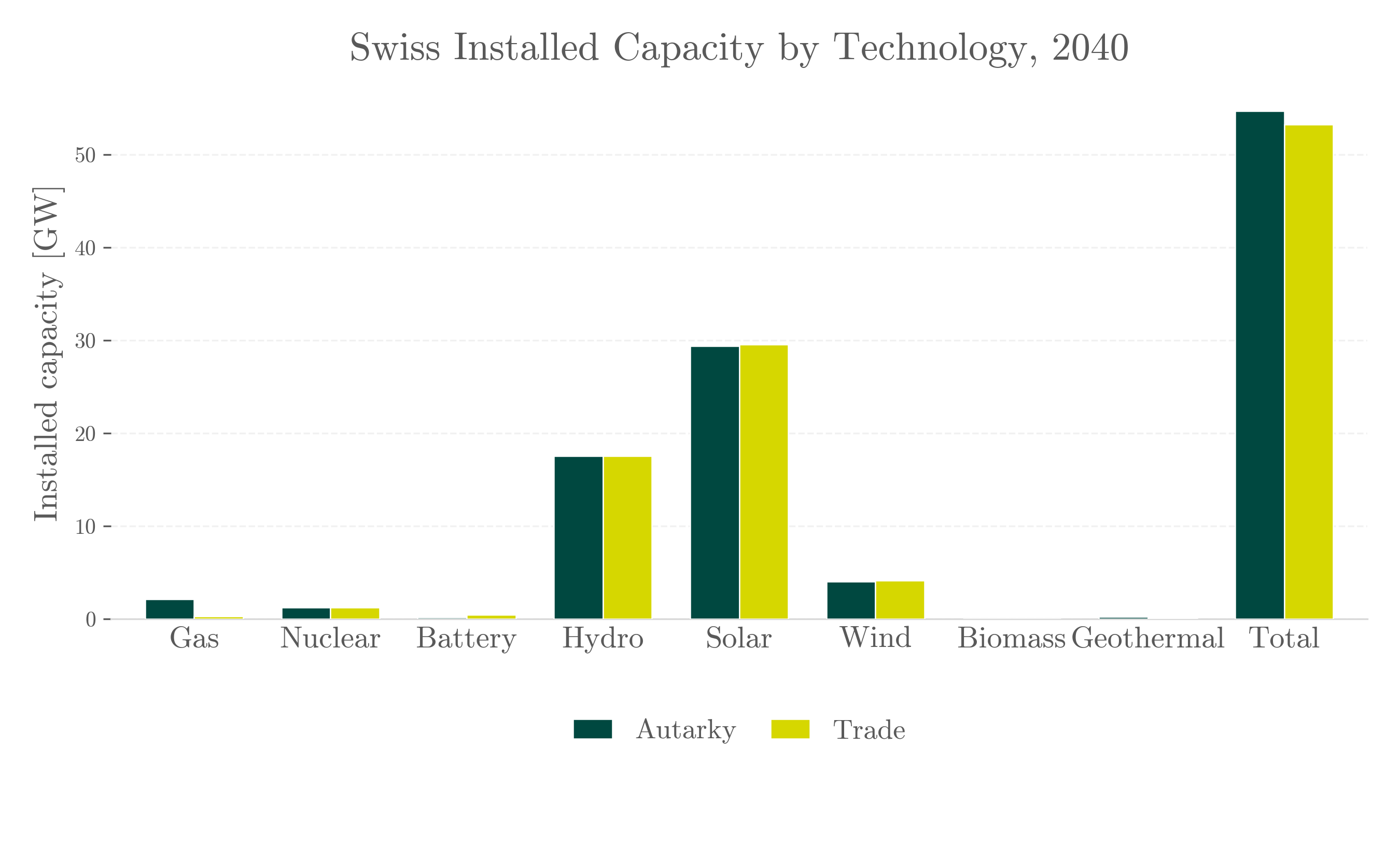}
    \begin{center}
    \caption{Swiss power generation mix.}
    \label{fig:generation_mix_absolute}
    \footnotesize
    \textit{Notes:} This figure shows the generation capacity required in 2040 under trade and autarky scenarios to guarantee a comparable security of supply. Most of the difference arises due to the backup gas capacity required in autarky. Autarky values relative to Trade for 2040 and absolute values for 2024 are shown in Figures \ref{fig:generation_mix} and \ref{fig:generation_mix_start}. 
    \end{center}
\end{figure}

Figure~\ref{fig:generation_mix_absolute} presents the generation mix for the investigated 2040 Autarky versus Trade scenarios and clearly shows a shift from renewables to gas as well as the additional required capacity.
The transition toward the 2040 electricity system is characterized by a decrease in nuclear capacity, an increase in electricity demand, a slight increase in hydropower capacity, and a renewable generation target.
To address these changes, CentIv identifies the optimal expansion under full market integration as the addition of 4.7 GW of wind, 22.9 GW of PV, and 0.3 GW of battery storage.
Under the assumption of an isolated electricity system in 2040, the optimal expansion instead entails 4.4 GW of wind, 22.7 GW of PV, 0.02 GW of battery storage, and 1.7 GW of gas generation.
In this scenario, internal flexible generation becomes more important than storage, hence the increase in gas generation and the reduction in battery storage.
This indicates that the additional battery storage in the integrated scenario is used to store excess renewable generation during the day and release it at night to export it to the neighboring countries.
These results are put in perspective with modeling results from other Swiss-focused studies in Appendix~\ref{app:other_results}.

\begin{figure}[H]
    \centering
    \begin{minipage}[c]{0.50\linewidth}
        \centering
        \includegraphics[width=\linewidth]{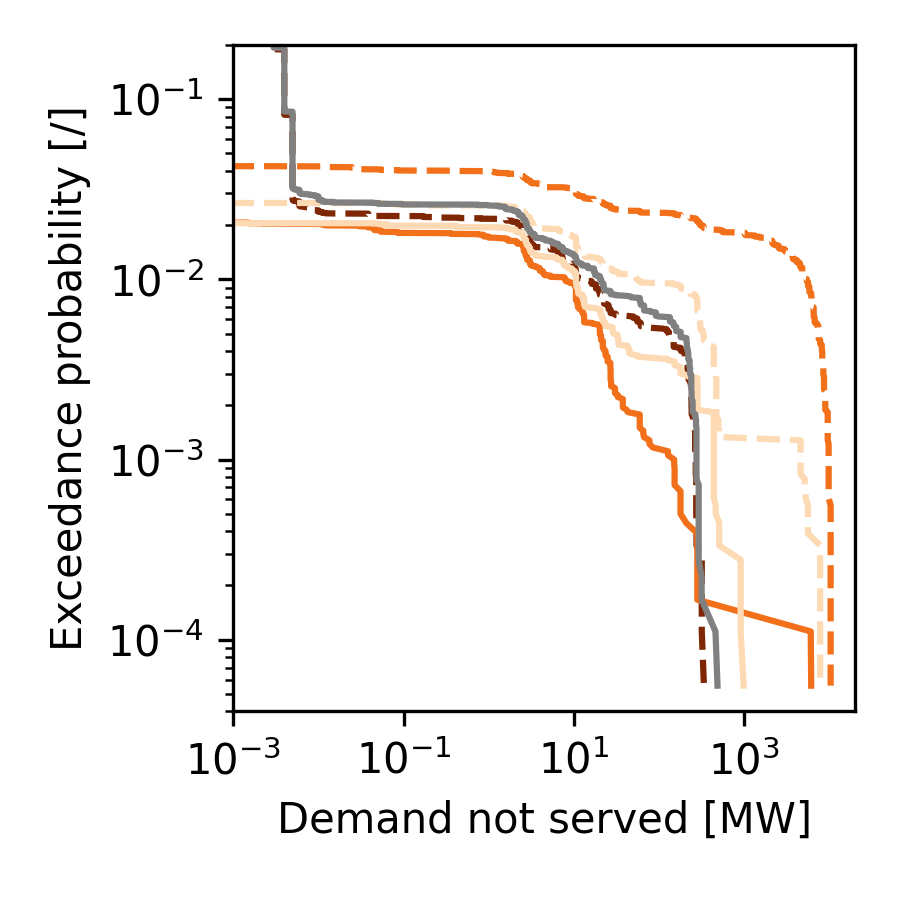}
    \end{minipage}\hfill
    \begin{minipage}[c]{0.50\linewidth}
        \centering
        \includegraphics[width=\linewidth]{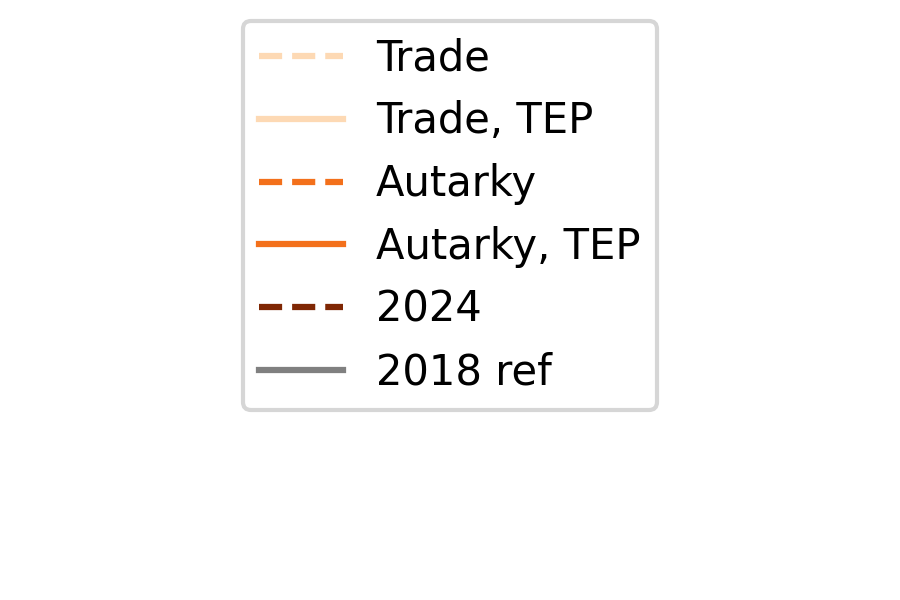}
    \end{minipage}
    \begin{center}
    \caption{Exceedance probability of demand not served (DNS)}
    \label{fig:cascades}
    \footnotesize
    \textit{Note:} This figure shows the exceedance probability of demand not served (DNS) for the Trade and Autarky scenarios, before and after transmission expansion planning (TEP), as well as for the 2024 and the 2018 ref scenarios. The total risk for the scenarios (the same order as the legend) is: 157, 19, 1435, 5, 18, and 21 MW.
    \end{center}
\end{figure}

Cascades is used to assess system security and to identify the transmission expansion that achieves system security comparable to the reference curve.
Figure~\ref{fig:cascades} shows the risk curves, reporting the exceedance probability of demand not served (DNS).
It highlights that system security worsens for the 2040 scenarios if no transmission expansion is performed.
However, upgrading the transmission grid has a positive impact on security, reducing the probability and the magnitude of DNS to values comparable to the reference case.
For 2040, the integrated and isolated scenarios require the expansion of 5 and 14 lines, respectively.

The investments in generation, storage, and transmission capacity result in the costs reported in Table~\ref{tab:annualized_investment_costs}.
The table reports the annualized investment costs for assets selected by the CentIv and Cascades models.
In addition to the 2040 investment costs, additional costs are associated with the fixed hydro capacity expansion (734 million EUR) and the fixed transmission grid expansion (198 million EUR). Because these investments are already planned and are assumed to occur under both scenarios, they cancel out in the welfare comparison and are therefore excluded from the scenario-specific investment differential. Throughout the remainder of the paper, all monetary values are expressed in constant 2024 EUR, unless otherwise stated.
\begin{table}[H]
    \centering
    \caption{Annualized investment costs [million 2024 EUR]}
    \label{tab:annualized_investment_costs}
    \footnotesize
    \begin{tabular}{lll} \toprule
        Scenario    & Investment type                 & Costs [million 2024 EUR]\\ \toprule
        Autarky     & Generation and storage          & 3142\\
                    & Transmission                    & 44\\
        Trade       & Generation and storage          & 3047\\
                    & Transmission                    & 21\\ \bottomrule
    \end{tabular}

\end{table}

\subsection{Power price forecasting model}

Once the additional generation and transmission capacities have been determined, we simulate hourly wholesale market outcomes using the LPP\footnote{LPP stands for \textit{\textbf{L}angfristige Strom\textbf{p}reis\textbf{p}rognose} (German), or in English, long-term price projections model.} model. The LPP framework is developed and maintained by BKW, while its optimization engine is supplied by THEMA Consulting. The model takes, among other inputs, an hourly demand vector, a detailed representation of the generation fleet, and a set of parameters describing the financial, technical, and climatic environment. On this basis, an optimization algorithm derives hourly bids for each power plant and constructs the corresponding hourly supply function, known as the merit-order curve, by ranking bids in ascending order.

The modeling environment is characterized by hourly weather conditions, temperature, solar irradiation, precipitation, and wind, which affect electricity demand (in particular through temperature) and renewable supply (through irradiation, precipitation, and wind). The financial environment is also represented and is described in Table \ref{tab:Fuels_prices_ref}; it includes fuel prices (gas, H2, and coal) as well as the availability and price of \coo\ allowances. Those prices are used in both Nexus-e and LPP models. Finally, cross-border net transfer capacities between countries and regions are explicitly specified and are consistent with the values in Table \ref{tab:ntc_ch}. Alternative fuel- and carbon-price assumptions are examined in Appendix \ref{app:robust}. Although the qualitative welfare ranking is preserved, the magnitude of the welfare gain varies materially.

\begin{table}[htbp] 
\centering 
\begin{threeparttable} 
\caption{Commodity and carbon price assumptions} 
\label{tab:Fuels_prices_ref} 
\begin{tabular}{lrrrrr} 
\toprule Scenarios & $CO_2$ & $H_2$ & Gas & Coal & Nuclear\\ 
\midrule 
Prices & 147 & 109.4 & 35.3 & 15.7 & 6.8 \\
\bottomrule 
\end{tabular} 
\begin{tablenotes}
\footnotesize 
\item 
\textit{Note:} CO\(_2\): 2024 EUR/tCO\(_2\); fuels: 2024 EUR/MWh.
\end{tablenotes}
\end{threeparttable} 
\end{table}

Each generation unit is described by a set of technical and economic parameters, including installed capacity, ramp-up and ramp-down limits and costs, and \coo\ emissions. For hydropower plants, reservoir constraints are incorporated as well, notably precipitation-dependent inflow profiles, turbine and pump capacities and storage volumes. Taken together, these features allow producers to determine profit-maximizing bidding strategies in the day-ahead market.

On the demand side, the framework simulates multiple demand categories with distinct load profiles and volumes. This enables a granular representation of consumption patterns, such as service-sector demand concentrated during working hours, relatively stable baseload demand from data centers, and late-afternoon demand from residential electric-vehicle charging. The long-term evolution of each demand segment is calculated using a combination of policy targets and subsidy schemes, published investment plans, technical reports, and macroeconomic forecasting models. Finally, the model solves an annual optimization problem comprising 8,736 interconnected hourly time steps, in which hourly generation and demand schedules are optimized to maximize total social welfare. This results in hourly power plant dispatch and price estimates.

\section{Economic Consequences}\label{sec:study}

Economic consequences of energy policies are numerous, heterogeneous, and transmitted through distinct channels. To capture this broad range of impacts while maintaining a simple and transparent framework, thereby ensuring traceability and interpretability, we adopt two complementary frameworks. We begin by isolating the implications for the electricity market in a deliberately stylized setting. To this end, we assume perfectly price-inelastic demand and hold economic activity fixed, thereby excluding any \textit{a priori} adjustment by consumers or producers. This ensures a clean comparison between trade and autarky, with hourly electricity demand—both its aggregate level and technological composition—held identical across scenarios throughout the simulation horizon.

To complement this electricity-market welfare analysis,
we consider the potential broader macroeconomic
consequences of persistent electricity-price differences
between the two scenarios. Higher electricity prices
under autarky may affect economic activity by reducing
the competitiveness of electricity-intensive production,
potentially inducing adjustments in output and firm
location. To explore these effects, we develop an
illustrative framework linking electricity demand
adjustments to economic activity through the historical
relationship between temperature-adjusted electricity
consumption and GDP. Unlike the welfare analysis,
this second approach allows for adjustments in
electricity demand but does not aim to identify a
causal effect on GDP. The two approaches therefore
capture distinct economic dimensions and should
not be interpreted as additive measures of the
cost of autarky.

\subsection{Surplus analysis}
We decompose the welfare impact of cross-border electricity trade into four components. First, under the assumption of perfectly inelastic demand, consumer surplus (CS), as defined in Eq. (1), is given by the difference in prices across the two scenarios multiplied by the quantity demanded and summed over all simulated hours and consumers. Second, the producer surplus (PS) can be directly estimated using outcomes of our fundamental model. More specifically, the producer surplus differential between the two scenarios is computed over all simulated hours and producers. Such values correspond to the market price minus marginal cost, multiplied by the quantity of power produced, and summed across producers and hours as shown in Eq. (2).  Finally, additional grid and generation costs represent the investment costs of all required new lines and plants, annualized considering the service life, including technology-specific financing costs as defined in Eqs. (3) and (4) respectively. We compute these components as follows:

\begin{equation}
    \Delta CS = CS^{T} - CS^{A} =  \sum_{i \in Cons} \sum_{h=1}^{H} \left[\left(p_{h}^A - p_{h}^{T}\right)\cdot Q_{i,h}^d \right]
\end{equation}

\begin{equation}
    \Delta PS = PS^{T} - PS^{A} = \sum_{j \in Prod} \sum_{h=1}^{H}  \left[ (p_{h}^{T} - MC_{j, h}) \cdot Q_{j,h}^{s,T} - (p_{h}^{A} - MC_{j, h}) \cdot Q_{j,h}^{s,A} \right]
\end{equation}

\begin{equation}
   \Delta Grid = \left(\sum_{l=1}^{L^A}F_l\frac{r_{l}}{1-(1+r_{l})^{-T_l}}\right)  - \left(\sum_{l=1}^{L^T}F_l\frac{r_{l}}{1-(1+r_{l})^{-T_l}}\right)
\end{equation}

\begin{equation}
    \Delta Gen = \left(\sum_{p=1}^{P^A}F_p\frac{r_{p}}{1-(1+r_{p})^{-T_p}}\right)  - \left(\sum_{p=1}^{P^T}F_p\frac{r_{p}}{1-(1+r_{p})^{-T_p}}\right) 
\end{equation}

where $i$ and $j$ index agents in the sets of consumers ($Cons$) and producers ($Prod$), respectively, and $h$ indexes the hours of the year 2040. Let $p_h^{A}$ and $p_h^{T}$ denote the day-ahead price in hour $h$ under the autarky and trade scenarios, respectively, while $MC_{j,h}$ is the marginal cost of producer $j$ in hour $h$. $Q_{i,h}^d$ and $Q_{j,h}^s$ denote the hourly electricity demand of consumer $i$ and the hourly electricity supply of producer $j$, respectively.\footnote{Note that while demanded quantity remains identical across scenarios, supplied quantity varies due to international trade changes.} Finally, $F_l$ and $F_p$ are the fixed costs associated with transmission grid expansion and with the $N$ additional power plants required under autarky, with $T_l$ and $T_p$ representing the lifetimes of lines and plants respectively. $r$ denotes the financing interest rate.

Let $\Delta$ denote the difference between the scenarios, such that $\Delta>0$ implies higher welfare under the trade scenario. The resulting total welfare ($TW$) effect is thus given by:

\begin{equation}
    \Delta TW = \Delta CS + \Delta PS + \Delta Grid + \Delta Gen.
\end{equation}

Proceeding in this way to estimate the economic benefits of international electricity trade for Switzerland has several advantages. First, the use of a fundamental model limits reliance on statistical estimation and provides a transparent and reproducible framework. Second, it allows for a comprehensive assessment of welfare effects within the Swiss electricity market, which would be the sector most directly affected by a potential shift towards autarky. However, this evaluation relies on the restrictive assumption of constant electricity demand, implying no response from economic agents to the substantial price changes induced by autarky. In practice, higher electricity prices would trigger adjustments in consumption and production, which may mitigate the initial price increases but also entail broader economic costs that are not captured by the power-market welfare analysis. The next section therefore relaxes the fixed-demand assumption and investigates the potential macroeconomic implications of such adjustments.

\subsection{Electricity-based macroeconomic mapping}

In the interest of simplicity and consistency, we develop an alternative methodology based on our fundamental power-market model and electricity-market data, avoiding the complexity of dynamic macroeconomic models. Rather than attempting to estimate the causal effect of electricity prices on macroeconomic variables, we exploit a property of electricity consumption extensively documented in the economic literature: its ability to serve as a proxy for economic activity and GDP.

Our approach builds on the distinction between the two scenarios considered in this study. The Trade scenario represents the continuation of Switzerland's current access to international electricity markets and serves as our baseline long-run economic equilibrium in 2040. Given the time horizon considered, we assume that firms have had sufficient time to adjust their production, investment, and location decisions to prevailing economic conditions. The Trade scenario therefore provides the reference level of economic activity against which the consequences of Autarky are assessed.

Autarky, by contrast, represents a deviation from this long-run equilibrium. By construction, all other relevant parameters, including the international economic environment, financial conditions, and security-of-supply requirements, among others, are held constant across scenarios. The only difference arises from the conditions governing electricity supply  and trade in Switzerland. As shown by our fundamental power-market model, this deviation results in substantially higher domestic electricity prices, thereby reducing the competitiveness of the Swiss economy, relative to the Trade benchmark. Therefore, the only driver of production and consumption decisions is the difference between Swiss and foreign electricity prices.

Higher electricity prices create incentives for industries to reduce production or relocate abroad. In addition, higher production costs increase prices, which reduce consumption of domestic goods and services. Such adjustments reduce domestic electricity demand and, consequently, exert downward pressure on Swiss electricity prices. This power-price adjustment mechanism operates through economic activity and not cross-border electricity arbitrage, absent under Autarky. Conversely, if domestic electricity prices were to fall below the Trade benchmark, Switzerland would become relatively more attractive, encouraging production and exerting upward pressure on electricity demand and prices.

Therefore, assuming that economic activity responds sufficiently strongly to electricity-price differences, Switzerland's relative electricity-price position vis-à-vis Europe converges across the two otherwise comparable counterfactual scenarios:  

\begin{equation}
\left(\frac{P_{t}^{CH}}{P_{t}^{EU}}\right)^{\mathrm{Trade}}
=
\left(\frac{P_{t}^{CH}}{P_{t}^{EU}}\right)^{\mathrm{Autarky}}.
\end{equation}

Since European electricity prices are assumed to be identical across both scenarios due to the relative market size, this condition translates into convergence of Swiss electricity prices across the Trade and Autarky cases.

Consequently, we iteratively decrease electricity consumption under Autarky in our fundamental model until the load-weighted average electricity price converges to that obtained under Trade. At each iteration, electricity demand is reduced by a uniform percentage across all hours, preserving the original hourly consumption profile. This ensures that the adjustment does not selectively target high-price hours to achieve price convergence. This procedure identifies the reduction in electricity demand required to restore the baseline price level, conditional on our long-run adjustment hypothesis. We then use the historical relationship between electricity consumption and GDP to map this adjustment into a GDP-equivalent effect.

Electricity consumption data for Switzerland are obtained from the Swiss transmission system operator, Swissgrid, and cover the period 2009–2025 at an hourly frequency. We voluntarily exclude prior data in order to avoid any bias arising from the evolution of electricity-related process technologies. In addition, such a range is comparable to the time until 2040. If the estimated relationship remains stable over the historical sample, this would support its application to 2040. Real quarterly GDP data are published by the State Secretariat for Economic Affairs (SECO).

Among the main determinants of electricity demand, temperature plays a major role and must therefore be controlled for. Since heating and cooling represent a significant share of power demand unrelated to economic activity, we follow \cite{FEZZI2021103907} and construct a temperature-adjusted electricity consumption series to isolate the relation between GDP and electricity consumption. We begin by estimating the relationship between electricity consumption and temperature as follows:

\begin{equation}
E_{t}
=
\beta_{0}
+\beta_{1}\,\overline{T}_{t}
+\beta_{2}\,\overline{T}_{t}^{2}
+\beta_{3}\left|\left(\overline{T}_{t}-T_{\text{sup}}\right) D^{\text{high}}_{t}\right|
+\beta_{4}\left|\left(\overline{T}_{t}-T_{\text{inf}}\right) D^{\text{low}}_{t}\right|
+\varepsilon_{t}
\end{equation}

where $E_{t}$ denotes electricity consumption in month $t$ and $\overline{T}_{t}$ denotes the average temperature in month $t$. The thresholds $T_{\text{inf}}$ and $T_{\text{sup}}$ define the lower and upper bounds of the “standard” temperature range and are set as $\overline{T}\pm\sigma_{T}$, with $\sigma_{T}$ denoting the standard deviation of the temperature. The indicators $D^{\text{low}}_{t}$ and $D^{\text{high}}_{t}$ take the value one when $\overline{T}_{t}$ falls below $T_{\text{inf}}$ or exceeds $T_{\text{sup}}$, respectively, and zero otherwise. When temperatures are unusually low (high), the baseline specification is augmented by a term capturing the magnitude of the deviation from the standard range. We then use the estimated coefficients to construct the temperature-induced component of electricity consumption, $\widehat{E}_{t}$.

\begin{figure}[h]
    \centering
    \includegraphics[width=1\linewidth]{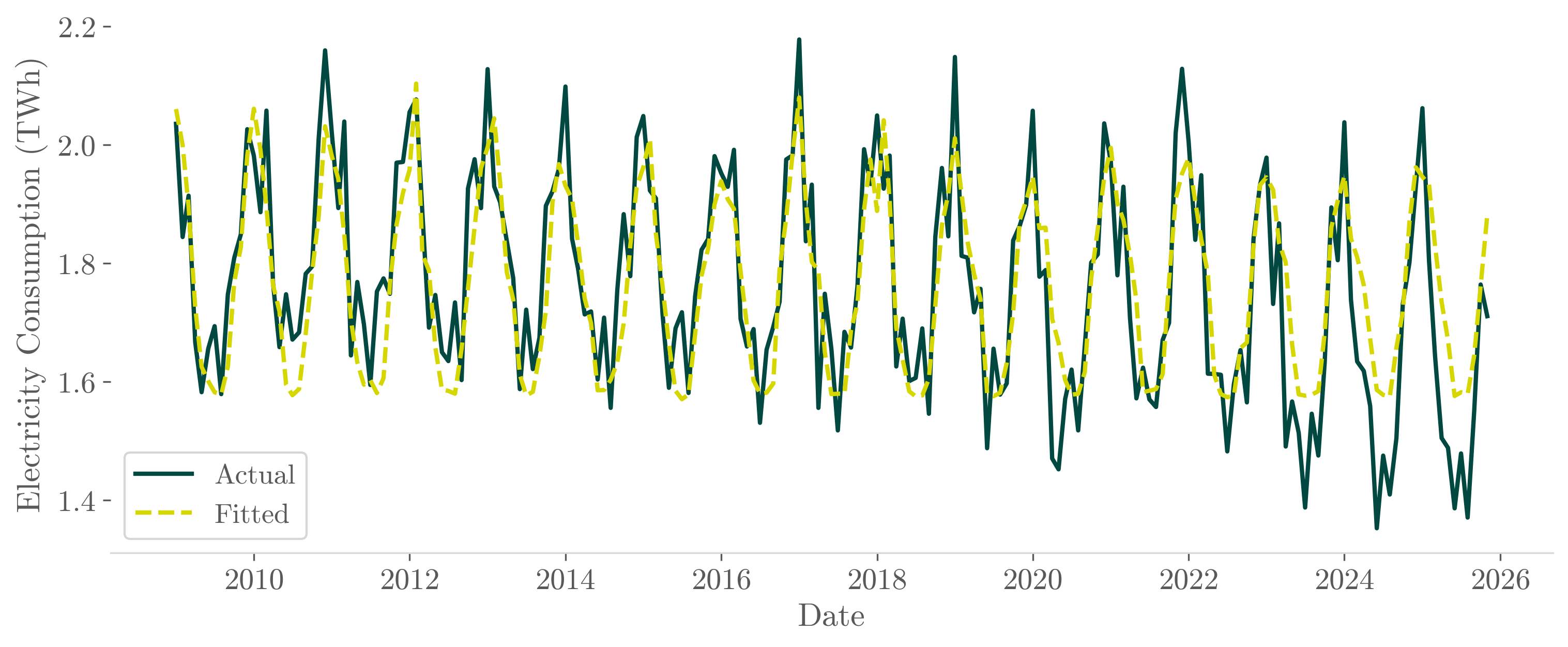}
    \begin{center}
    \caption{Estimation of electricity consumption based on temperature.}
    \label{fig:temp_elec_cons}
    \footnotesize
    \textit{Note:} This figure shows the actual and temperature-based estimated monthly consumption of electricity.  
    \end{center}
    
\end{figure}

Figure ~\ref{fig:temp_elec_cons} provides several important insights. A substantial share of electricity consumption is driven, directly or indirectly, by temperature. Both high and low temperatures increase demand for cooling and heating. Moreover, lower temperatures tend to coincide with darker periods (winter months and, more generally, shorter daylight hours), which raises the need for artificial lighting and further increases electricity consumption.

This allows us to construct a temperature-adjusted electricity consumption series. We then extract the component of electricity consumption that is orthogonal to temperature—i.e., the difference between observed and fitted values in Figure ~\ref{fig:temp_elec_cons}—and rescale it using the sample mean of electricity consumption as follows:

\begin{equation}
E^{\text{adj}}_{t}
=
E_{t}
-\widehat{E}_{t}
+\overline{E}
\end{equation}

The resulting monthly temperature-adjusted electricity consumption series is then aggregated to quarterly frequency by summing the monthly values, ensuring consistency with the frequency of GDP data. This procedure yields a series that is strongly correlated with GDP, with a Spearman correlation coefficient of $0.495$ based on year-over-year quarterly changes.\footnote{Moreover, the estimated values $\widehat{E}_{t}$ exhibit an approximately zero correlation with GDP ($-0.031$), consistent with the interpretation that they primarily capture temperature-driven variation rather than information about economic activity.} 

Finally, we estimate the empirical association between four-quarter changes in temperature-adjusted electricity consumption and four-quarter changes in real GDP. The baseline specification is:

\begin{equation}
g^{GDP}_{t}
=
\alpha
+\beta g^{E^{adj}}_{t}
+\sum_{q=2}^{4}\gamma_q D_{q,t}
+\varepsilon_t,
\label{eq:gdp_electricity}
\end{equation}

where $g^{GDP}_{t}$ denotes the four-quarter log change in real Swiss GDP and $g^{E^{adj}}_{t}$ the corresponding four-quarter log change in temperature-adjusted electricity consumption. $D_{q,t}$ is a quarter dummy equal to $1$ if observation $t$ belongs to quarter $q$ and $0$ otherwise, with the first quarter omitted as the reference category.

The baseline regression is estimated only on observations whose four-quarter comparison window does not overlap either the COVID-19 period or the European energy crisis. Because the dependent and explanatory variables are defined as four-quarter changes, an observation can remain affected by an exceptional episode even after the contemporaneous crisis period has ended. We therefore exclude any observation for which the interval $[t-4,t]$ overlaps COVID-19 (2020Q1--2021Q4) or the European energy crisis (2022Q1--2023Q4). This strict sample is intended to recover the historical GDP--electricity relationship under comparatively normal macroeconomic conditions.

The parameter $\beta$ measures the historical association between electricity-consumption growth and GDP growth in normal times. Electricity consumption is not interpreted as a causal production input in this exercise. Instead, temperature-adjusted electricity demand is used as a proxy for the scale of aggregate economic adjustment. Accordingly, $\beta$ is interpreted as an elasticity-like historical mapping rather than as a causal structural parameter. The quarter fixed effects $\gamma_q$ absorb systematic seasonal differences, $\alpha$ is the intercept, and $\varepsilon_t$ is the error term.

Because four-quarter growth rates generate overlapping observations and potentially reduce the effective information content of the sample, inference relies on heteroskedasticity- and autocorrelation-consistent Newey--West standard errors with four lags and a small-sample correction. To further assess the sensitivity of our results to overlapping observations, we estimate an alternative specification using non-overlapping Q4-to-Q4 annual observations. This specification retains one observation per year, eliminating the mechanical overlap between consecutive growth windows. Results are reported in Table~\ref{tab:overlap_robustness} in Appendix~\ref{app:robust_gdp}.

To assess whether the normal-times mapping is transportable to exceptional macroeconomic environments, we also estimate the full-sample specification

\begin{equation}
g^{GDP}_{t}
=
\alpha
+\beta g^{E^{adj}}_{t}
+\sum_{q=2}^{4}\gamma_q D_{q,t}
+\delta_C C_t
+\delta_E X_t
+\phi_C \left(C_t\times g^{E^{adj}}_{t}\right)
+\phi_E \left(X_t \times g^{E^{adj}}_{t}\right)
+\varepsilon_t,
\label{eq:gdp_transportability}
\end{equation}

where $C_t$ and $X_t$ denote mutually exclusive four-quarter-window indicators for COVID-like and energy-crisis-like periods, respectively. The interaction coefficients $\phi_C$ and $\phi_E$ test whether the GDP--electricity relation differs from its normal-times value during these exceptional episodes. This specification is used as a historical transportability test rather than to generate alternative GDP-loss scenarios.

\section{Results}\label{sec:results}

\subsection{LPP wholesale market outcomes}

The LPP model produces a wide range of highly granular hourly outputs, from reservoir filling levels to renewable capture rates. However, our analysis requires only a limited subset of these variables. The key output is the simulated electricity price, which forms the basis of our analysis. Specifically, we use the simulated day-ahead electricity prices. While these prices differ from the tariffs paid by end consumers, particularly households, because they exclude network charges and taxes, they allow us to accurately isolate the contribution of the wholesale electricity price component.\footnote{In the first part of the analysis we assume identical demand across scenarios. Under this assumption, any network charges that are purely volumetric (i.e., defined as a fixed amount per kWh) would cancel out when comparing autarky and trade. Percentage-based taxes (e.g., VAT) would mechanically scale price differences; accounting for them would therefore tend to amplify, rather than overturn, the implied gap between the autarky and trade scenarios.}

\begin{figure}[H]
    \centering
    \includegraphics[
        width=1\linewidth,
        trim=0 0 0 1.15cm,
        clip
    ]{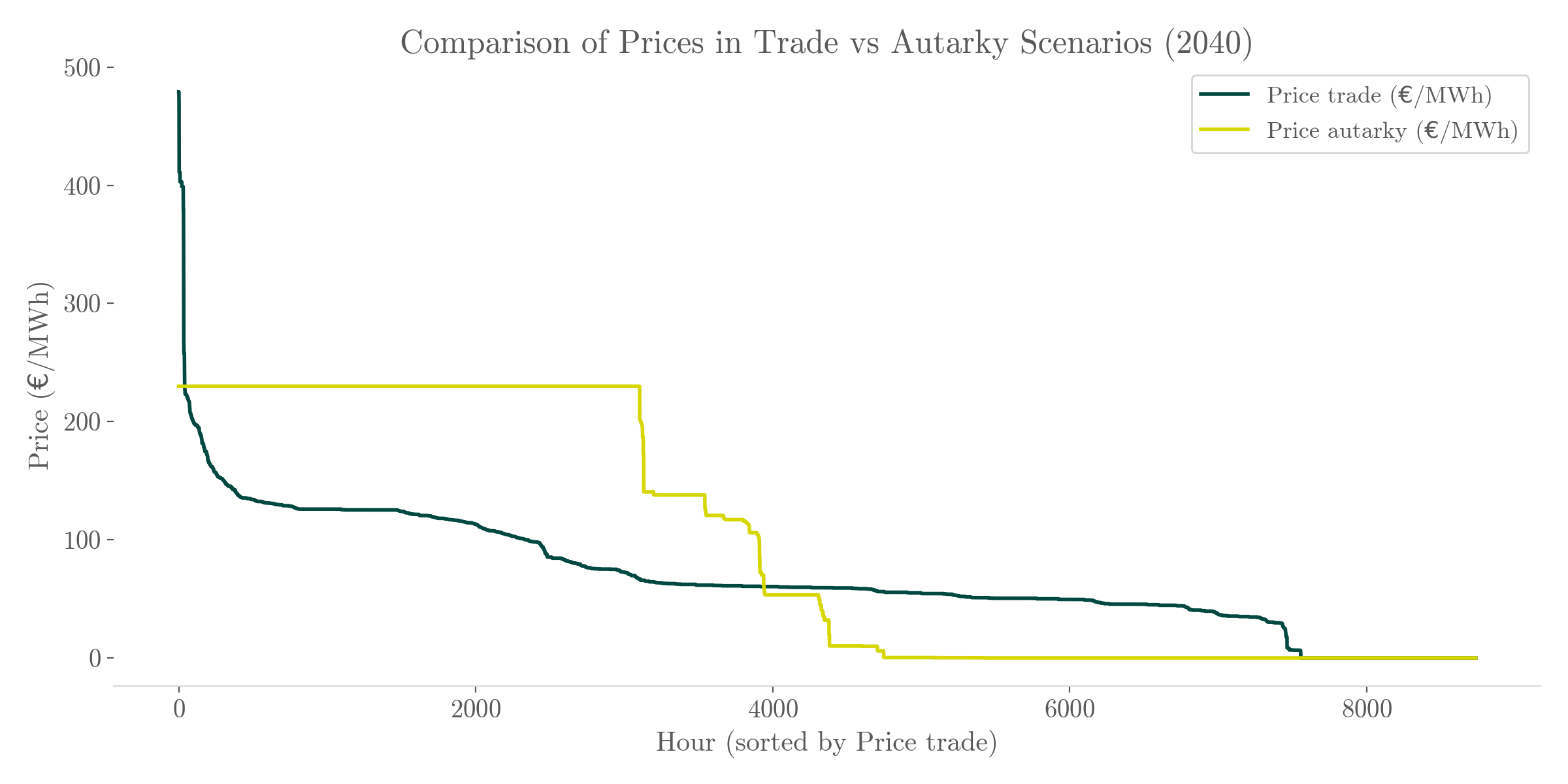}
    \begin{center}
    \caption{Price distribution.}
    \label{fig:Prices_repartition}
    \footnotesize
    \textit{Notes:} This figure shows the hourly price distribution for the two scenarios considered. Plotted series are computed by sorting all hourly prices in decreasing order. Monetary values are expressed in constant 2024 EUR.  
    \end{center}

\end{figure}

The first insight from Figure~\ref{fig:Prices_repartition} is that no scenario unambiguously dominates the other in terms of prices. On the one hand, because Switzerland has a comparatively high share of renewable generation capacity relative to many European countries, an autarky setting can yield lower prices in some periods by isolating the domestic market from higher prices in neighboring systems. In addition, the expansion of gas-fired capacity to secure supply during peak hours mitigates reliance on high-cost peaking technologies and thereby limits extreme price spikes in a small number of scarcity hours. On the other hand, these advantages are more than offset during substantial parts of the year by the unavailability of stable nuclear generation from France and low-cost wind generation from Germany, for instance, which, under cross-border trade, would otherwise contribute to lower prices in Switzerland. The  upper bound of prices, around 230 EUR/MWh and observed during a significant number of hours, is set by the most expensive generation technology installed in Switzerland (\textit{i.e.}, gas-fired generation). Such behavior is a direct consequence of marginal-cost pricing and autarky. Figure~\ref{fig:Prices_repartition} also demonstrates an interesting fact about hydropower reservoir capacities: the number of hours at maximum price together with the higher occurrence of zero prices show that the planned capacity is insufficient to arbitrage this in the autarky case. However, due to the moderate forecast horizon, any expansion beyond the projects already planned and considered in our study is not realistic.

\begin{figure}[H]
    \centering
    \begin{subfigure}[t]{0.49\linewidth}
        \centering
        \includegraphics[
        width=1\linewidth,
        trim=0 0 0 0.8cm,
        clip
        ]
        {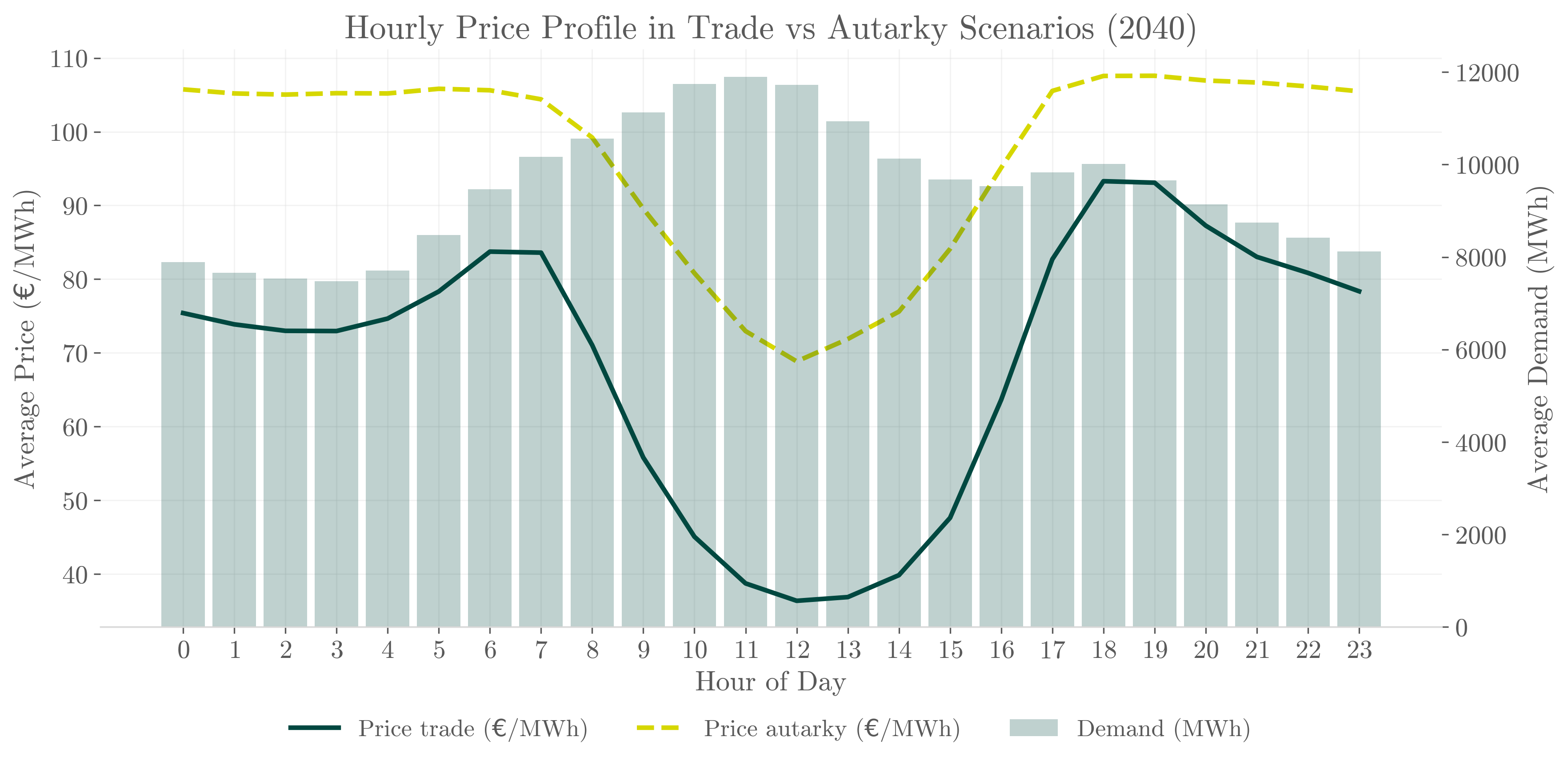}
        \caption{Hourly.}
        \label{fig:profiles_hourly}
    \end{subfigure}\hfill
    \begin{subfigure}[t]{0.49\linewidth}
        \centering
        \includegraphics[
        width=1\linewidth,
        trim=0 0 0 0.8cm,
        clip
        ]
        {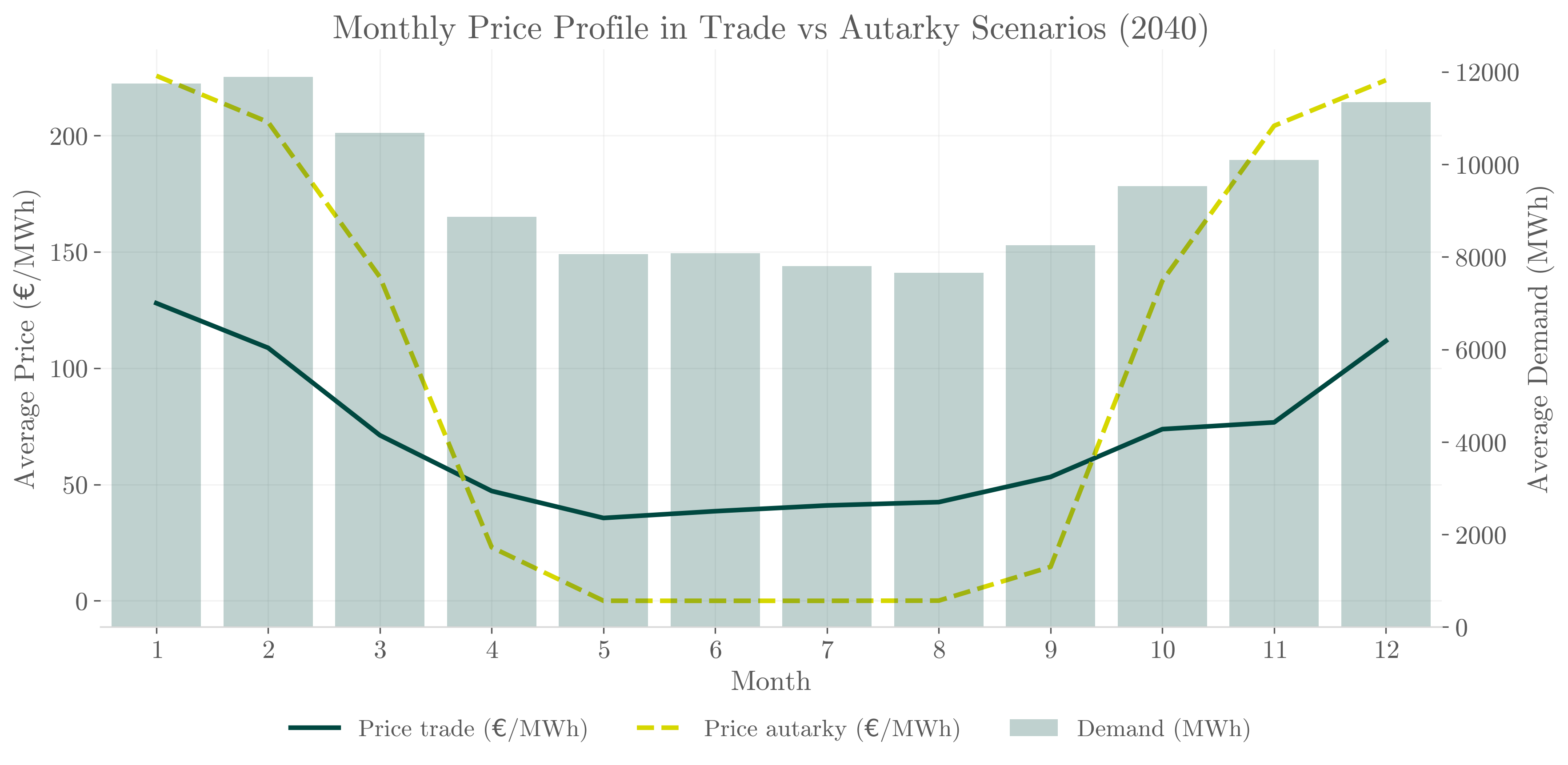}
        \caption{Monthly.}
        \label{fig:profiles_monthly}
    \end{subfigure}
    \begin{center}
    \caption{Electricity price profiles under trade and autarky.}
    \label{fig:profiles6}
    \footnotesize
    \textit{Notes:} This figure shows the price profiles for the two scenarios considered. The plotted series are obtained by computing the average prices over the corresponding periods (hours of the day or month) over the year 2040. The bars show the corresponding consumption for all periods. Monetary values are expressed in constant 2024 EUR.  
    \end{center}
\end{figure}

Additional insights into market conditions are provided by Figures ~\ref{fig:profiles_hourly} and \ref{fig:profiles_monthly}. These figures report hourly and monthly profiles for 2040, constructed by averaging prices and quantities across all observations belonging to a given hour of the day (or month of the year).

At the hourly level, average prices are 13 to 35 EUR/MWh higher under autarky than under trade. The price gap is the largest around noon, which coincides with the daily demand peak, suggesting potentially sizable implications for consumers. Turning to monthly averages, a similar pattern emerges: the benefits of trade are greatest during high-demand periods. However, the monthly profiles also indicate that autarky can be advantageous for consumers only during the summer months. This is consistent with Figure~\ref{fig:Prices_repartition} and reflects the combination of high solar irradiation, abundant baseload hydropower, and lower demand, which allows Switzerland to rely largely on renewable generation (\textit{i.e.}, generation with a marginal cost close to zero). However, lower prices during periods of high solar generation reduce incentives for PV investment, ultimately slowing future renewable energy deployment.

To obtain a clearer overview of the consequences, it is also informative to consider average prices over the year. In particular, the load-weighted annual average price provides a useful summary measure of the cost of electricity for consumers facing tariffs that are set on an annual basis and therefore reflect average procurement conditions, notably households. Moreover, an increasing share of demand is associated with consumers that operate continuously, such as data centers and uninterrupted industrial processes. This segment of demand, often referred to as inflexible load, is likewise primarily affected by average price levels rather than by short-run price fluctuations. Taken together, these considerations suggest that the aggregate price level can be at least as informative as the hourly profile for assessing the overall implications of the scenarios. By contrast, large consumers exposed to market-based contracts are more directly affected by hourly price volatility.

\begin{figure}[H]
    \begin{subfigure}[t]{1\linewidth}
    \centering
    \includegraphics[
        width=0.8\linewidth,
        trim=3 0 0 0cm,
        clip
        ]{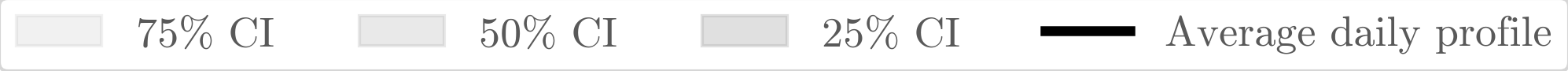}
    \end{subfigure}
    \centering
    \begin{subfigure}[t]{0.5\linewidth}
        \centering
        \includegraphics[
        width=1\linewidth,
        trim=0 0 0 0.8cm,
        clip
        ]
        {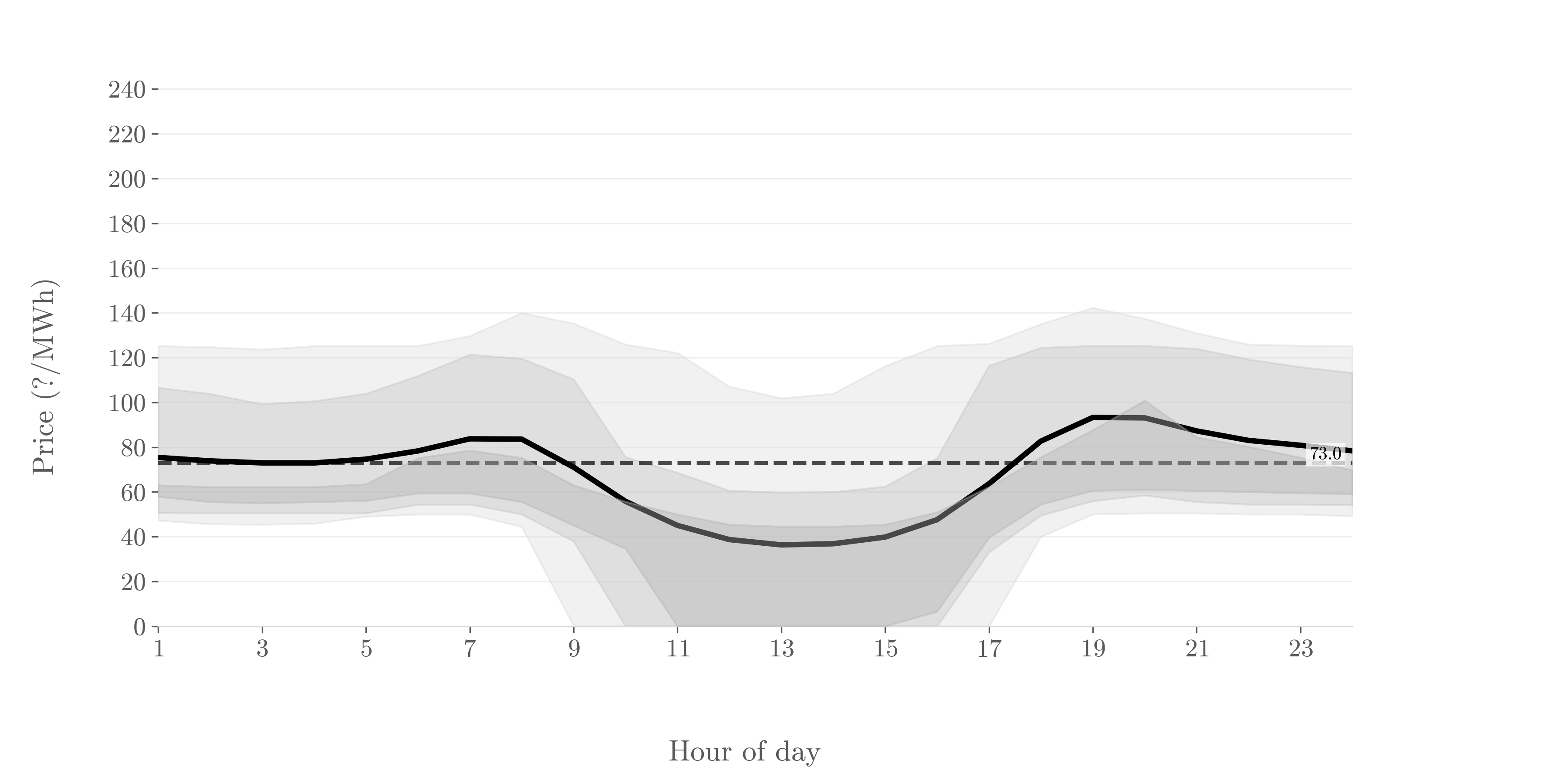}
        \caption{Trade.}
        \label{fig:profiles_hourly_spread_trade7}
    \end{subfigure}\hfill
    \begin{subfigure}[t]{0.49\linewidth}
        \centering
        \includegraphics[
        width=1\linewidth,
        trim=0 0 0 0.8cm,
        clip
        ]
        {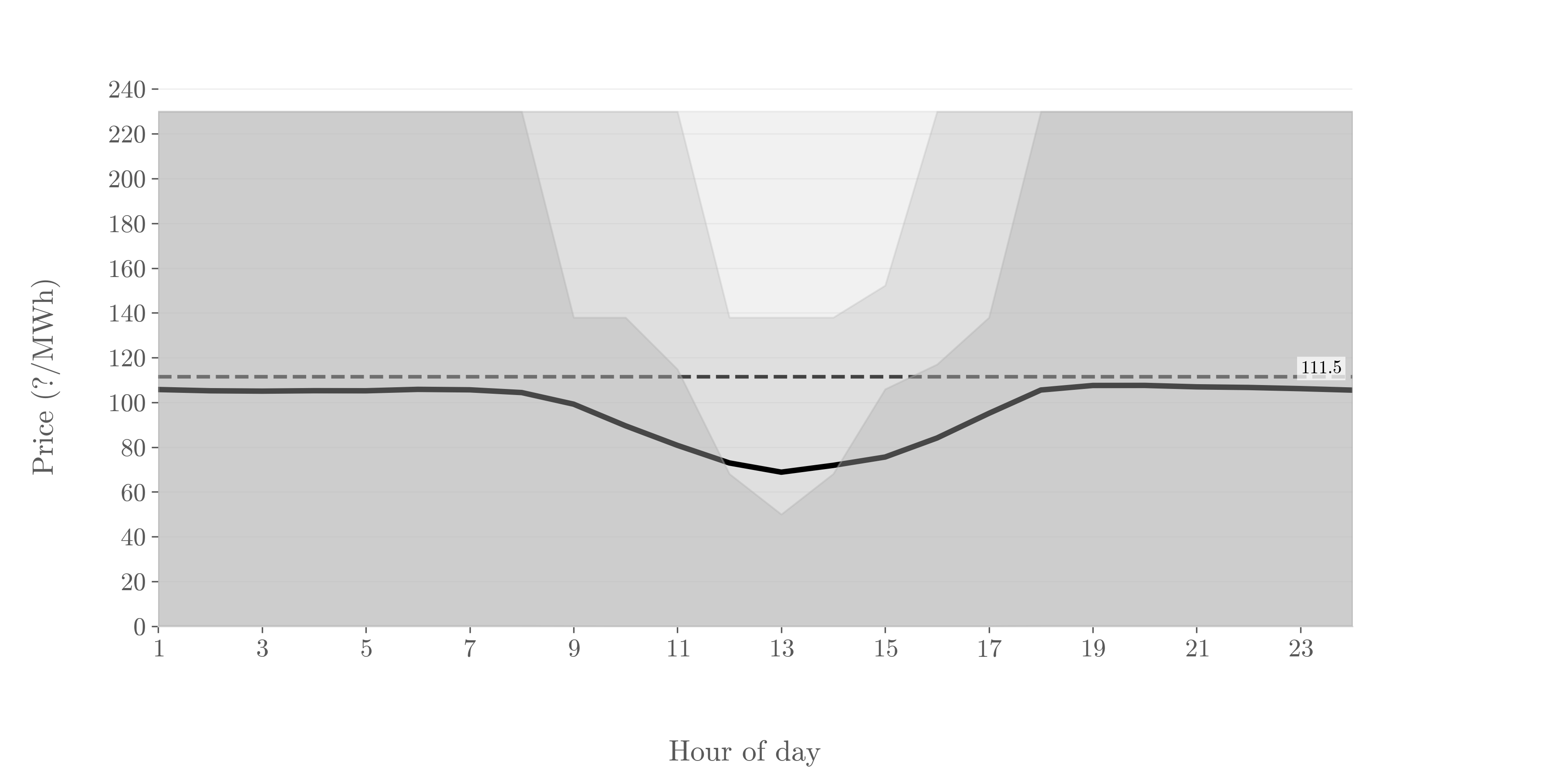}
        \caption{Autarky.}
        \label{fig:profiles_hourly_spread_autarky}
    \end{subfigure}
    \begin{center}
    \caption{Hourly electricity price profiles.}
    \label{fig:profiles_spread7}
    \footnotesize
    \textit{Notes:} This figure shows the hourly price profiles for the two scenarios considered. The plain lines are obtained by computing the average prices over the corresponding periods over the year 2040. The grey areas represent the volatility of the observations, showing within what range  25\%, 50\% and 75\% of prices fall. The horizontal dashed lines represent the yearly averages. Monetary values are expressed in constant 2024 EUR. 
    \end{center}
\end{figure}

While the solid lines in Figure~\ref{fig:profiles_spread7} reproduce the hourly profiles shown in Figure~\ref{fig:profiles6}, the additional elements allow us to draw two key conclusions regarding the impact of autarky on the electricity market: \textit{i)} the load-weighted annual average price increases by 53\% (from 73 to 111.5 EUR/MWh), suggesting a substantial loss in consumer surplus; and \textit{ii)} price volatility is higher, implying greater uncertainty.

\subsection{Welfare analysis}

\begin{figure}[H]
    \centering
    \includegraphics[
        width=0.9\linewidth,
        trim=0 0 0 1.15cm,
        clip
    ]{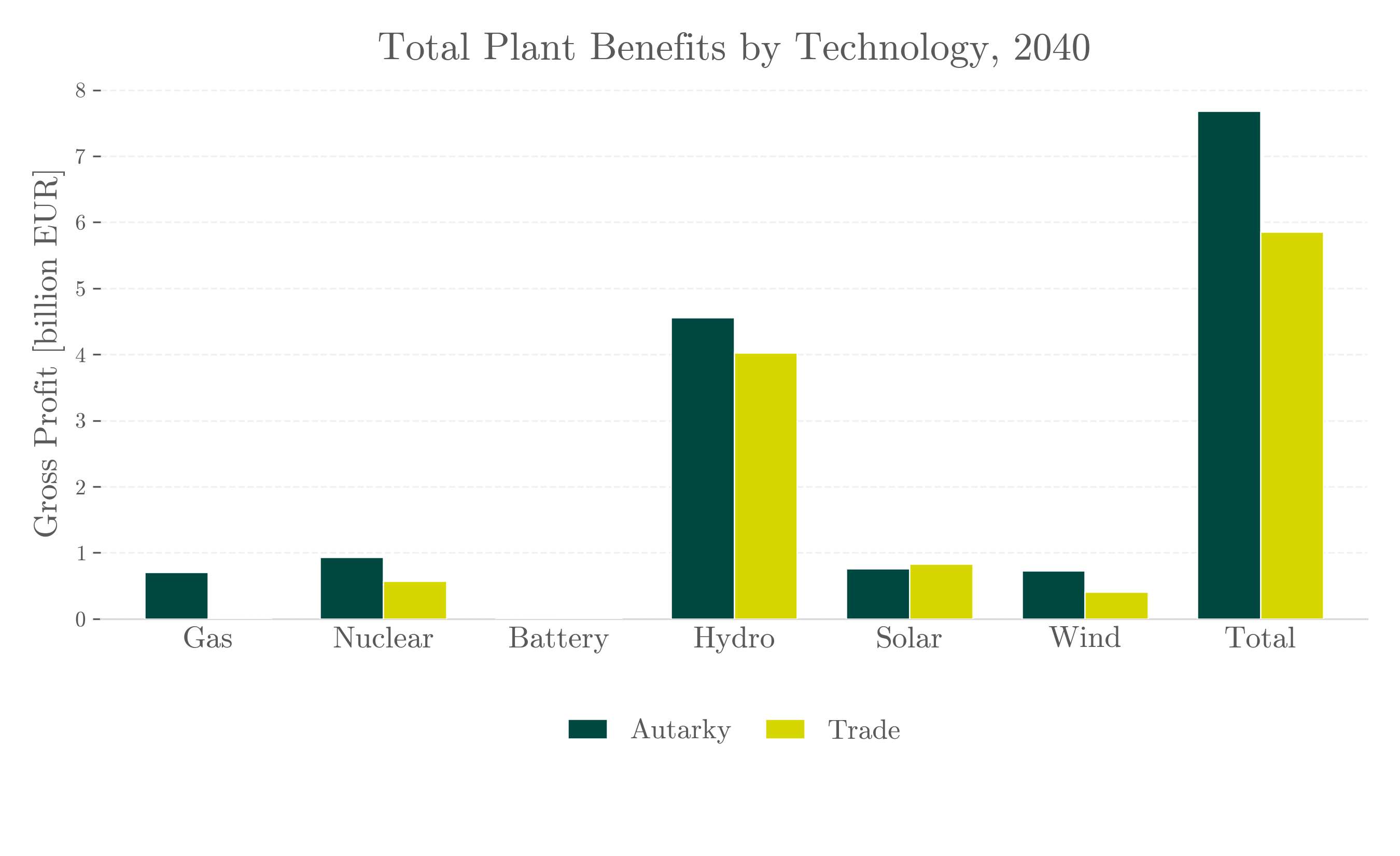}
    \begin{center}
    \caption{Technology-wise Gross Profit by scenario.}
    \label{fig:ProducerSurplus}
    \footnotesize
    \textit{Notes:} This figure shows Gross Profit for the different technologies used under the two scenarios. Waste is marginal, thus not considered in this plot. Monetary values are expressed in constant 2024 EUR.
    \end{center}
\end{figure}

As suggested by the preceding analysis and shown in Figure \ref{fig:ProducerSurplus}, producers benefit in general from the autarky scenario. Restricted imports of affordable electricity lead to higher average prices in Switzerland, which eventually increases producer surplus. With the exception of solar, all other power sources experience an increase in net revenues under Autarky, with gas seeing the most significant increase. While providing interesting insights, such data provide only partial information and do not reveal the mechanisms behind those results. 

\begin{figure}[H]
    \centering
    \includegraphics[
        width=0.9\linewidth,
        trim=0 0 0 1.15cm,
        clip
    ]{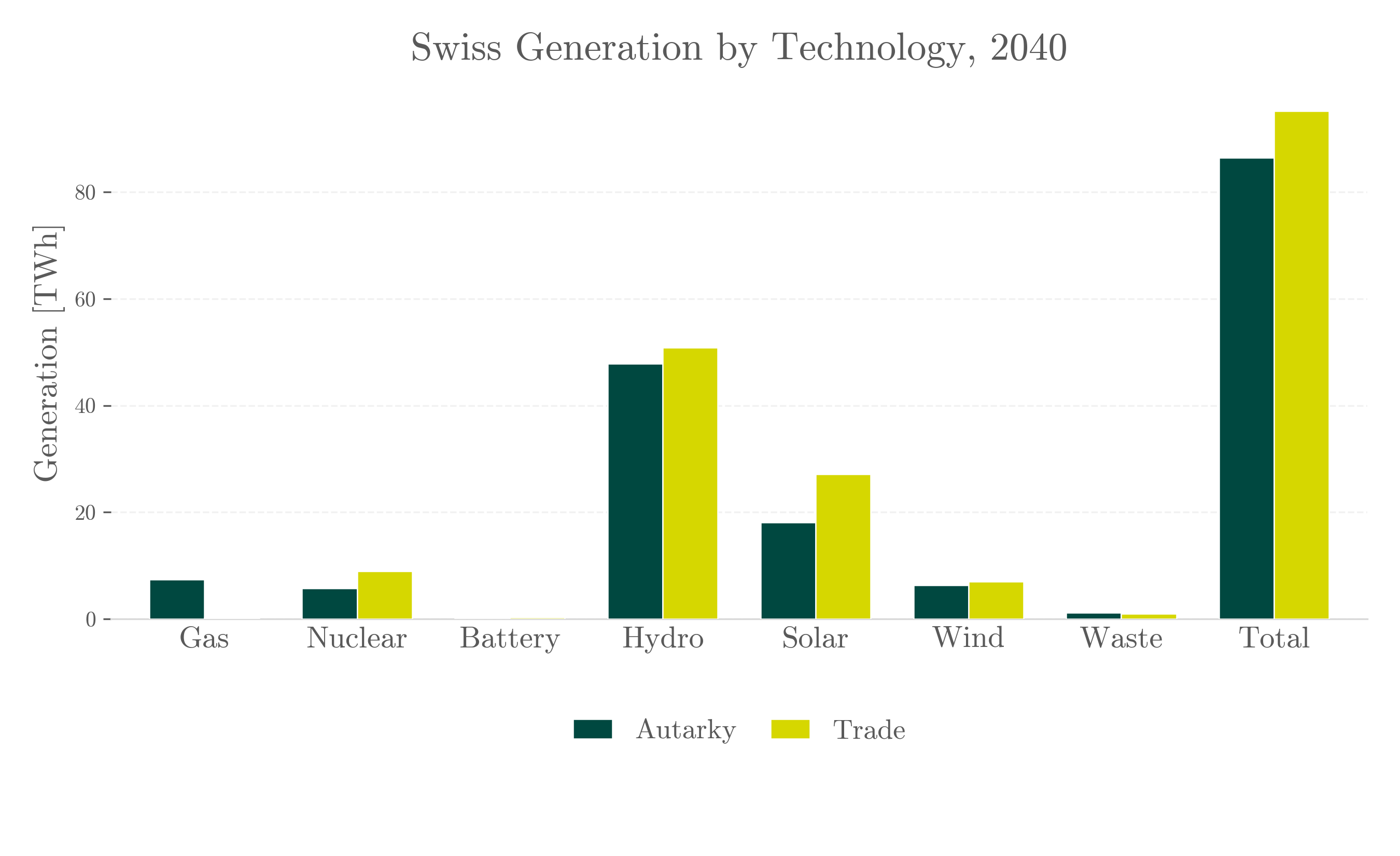}
    \begin{center}
    \caption{Technology-wise 2040 Generation in TWh.}
    \label{fig:ProducerGeneration}
    \footnotesize
    \textit{Notes:} This figure shows generation for the different technologies used under the two scenarios. 
    \end{center}
\end{figure}

The higher benefits for gas generators actually arise from the increased production. As mentioned in the introduction, cross-border electricity trade provides access to cheaper, often renewable, power when required. Under Autarky, insufficient domestic low-carbon generation must more frequently be supplemented by gas capacity to satisfy demand, leading to a significantly higher yearly generation as shown in Figure \ref{fig:ProducerGeneration}. In addition, the gas generation fleet is composed of multiple technologies associated with different marginal costs. Since under Trade the most costly plant is rarely activated, this provides fewer opportunities for the most efficient plants to avoid setting the market price, thus allowing for more profits.\footnote{This paper focuses on the day-ahead market only, neglecting revenues from other electricity-market segments such as capacity mechanisms, intraday market, balancing and ancillary services, important sources of profit for flexible generation plants.} Regarding PV, Figure \ref{fig:ProducerGeneration} provides an explanation for the lower surplus. The inability to export power under Autarky mechanically leads to higher curtailment and lower production, decreasing the final net revenues. Finally, it is worth mentioning that Autarky increases the net revenues per MWh for all generation technologies as shown in Figure \ref{fig:ProducerBenefitsMWh}.

According to our simulations, the trade scenario generates annual losses of approximately EUR~1.8~billion for producers. By contrast, the effect for other stakeholders is positive. Consumers, facing lower average prices, benefit by approximately EUR~3.2~billion per year, which more than offsets producers’ losses. Furthermore, while not priced in consumer surplus, the lower uncertainty would certainly represent an additional benefit for consumers. Considered together, results suggest a EUR 1.8 billion transfer from largely publicly owned producers to consumers. Beyond these direct effects, autarky also entails additional costs to maintain an acceptable level of security of supply, notably through generation and grid expansion. These costs depend on the assumed financing rate but, with annual losses of EUR 95 million and EUR 23 million, together equivalent to approximately 3.7\% of consumer surplus change, different interest rates impact the total results only marginally. In addition, results are shown to be robust to changes in fuel prices as shown in Table \ref{tab:economic-benefit-analysis_sensitivity}.

\begin{table}[H]
\centering

\caption{Economic Benefit Analysis (Million 2024 EUR)}
\begin{tabular}{rrrrrr}
\toprule
Year & $\Delta$ CS & $\Delta$ PS & $\Delta$ INV & $\Delta$ GRID & Total Welfare \\
\midrule
2040 & 3189 & -1826 & 95 & 23 & 1481 \\
\bottomrule
\end{tabular}\\
\footnotesize
\textit{Note:} Positive values indicate a welfare gain of Trade relative to Autarky.
\end{table}

Although some producers may benefit from higher prices, consumer losses, together with any additional investment required to secure supply, more than offset these gains, ultimately reducing total welfare. Overall, the welfare cost of autarky in the Swiss electricity market amounts to about EUR~1.5~billion in 2040. This amount corresponds to EUR 375 per Swiss household annually and about EUR 800 for consumer losses.

\subsection{GDP Mapping}

Following the framework described in Section \ref{sec:study}, and based on the literature-supported assumption that higher electricity prices affect production through lower activity and relocation, the first step is to identify the consumption level leading to similar prices in 2040 under both scenarios. Given that everything else is held constant, we make the hypothesis that the decrease in production should effectively stop when prices converge. For simplicity and given that GDP data are not available on an hourly basis, we consider that prices converge once the load-weighted annual average price is equalized under trade and autarky.

\begin{figure}[H]        
    \begin{subfigure}[t]{1\linewidth}
    \centering
        \includegraphics[
        width=0.8\linewidth,
        trim=3 0 0 0cm,
        clip
        ]{legend_profile.png}
    \end{subfigure}
    \begin{subfigure}[t]{0.49\linewidth}
        \centering
        \includegraphics[width=\linewidth]{price_profile_comparison_Price_trade_weighted.png}
        \caption{Hourly price profile (Trade).}
        \label{fig:profiles_hourly_spread_trade}
    \end{subfigure}\hfill
    \begin{subfigure}[t]{0.49\linewidth}
        \centering
        \includegraphics[width=\linewidth]{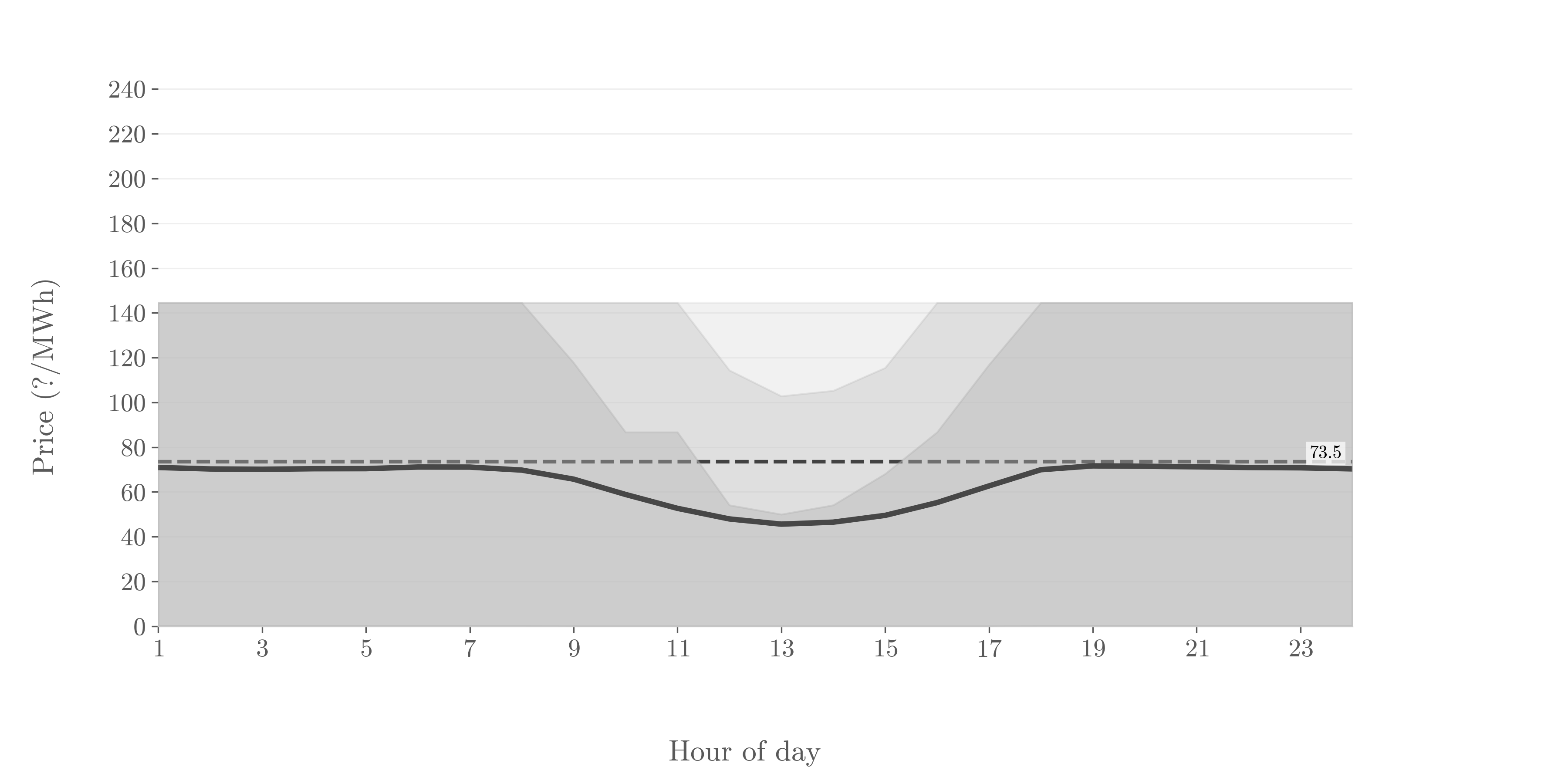}
        \caption{Hourly price profile (Autarky).}
        \label{fig:profiles_hourly_spread_autarky}
    \end{subfigure}
    \begin{center}
    \caption{Hourly electricity price profiles with lower demand.}
    \label{fig:profiles_spread}
    \footnotesize
    \textit{Notes:} This figure shows the hourly price profiles for the two scenarios considered with a lowered autarky demand leading to the convergence of average prices. The plain lines are obtained by computing the average prices over the corresponding periods over the year 2040. The grey areas represent the volatility of the observations, showing within what range  25\%, 50\% and 75\% of prices fall. The horizontal dashed lines represent the yearly averages. Monetary values are expressed in constant 2024 EUR.  
    \end{center}
\end{figure}

The required reduction corresponds to 1.14~TWh as shown in Figure \ref{fig:profiles_spread}. At first glance, this may appear small relative to the projected total consumption of about 80~TWh in 2040. However, it should be interpreted in light of the evolution of other demand components. In our framework, an autarky policy does not alter environmental targets for consumption: the objective of net-zero emissions by 2050 remains unchanged. Consequently, demand associated with electric vehicles, heat pumps, and electrolyzers can be treated as exogenous. Table~\ref{tab:demand_evolution} summarizes these developments and the implied differences. 

\begin{table}[H]
\centering
\caption{Evolution of demand by category (in TWh)}
\label{tab:demand_evolution}
\begin{tabular}{lccc}
\hline
\textbf{Category} & \textbf{2026} & \textbf{2040} & \textbf{Diff} \\
\hline
EV      & 1.46 & 13.86 & 12.40 \\
P2X     & 0.11 & 1.00  & 0.89  \\
Cooling & 0.84 & 1.14  & 0.30  \\
Heating & 3.73 & 8.00  & 4.27  \\
\hline
\textbf{Total} & \textbf{6.14} & \textbf{24.00} &  \textbf{17.86} \\
\hline
\end{tabular}
\end{table}

In the baseline calculation, these demand components are treated as policy- or technology-driven and therefore held constant across scenarios. According to  Table \ref{tab:demand_evolution}, this can be achieved by subtracting the additional 17.9 TWh from the total 80 TWh. This 1.14 TWh reduction has to be considered in light of the resulting 62.1 TWh of traditional demand, corresponding to an estimated decrease of 1.8\%.

Finally, to derive a GDP proxy based on electricity consumption, we estimate the relationship between these two variables as defined in Eq. (9). Since the period covered by our data includes several significant crises, such as the COVID-19 pandemic and the energy crisis following Russia’s aggression in Ukraine, we retain data from normal periods, up to 2019Q4, in our main specification, while later periods are used for robustness tests. Table \ref{tab:gdp_main_mapping} reports the main results and shows significant co-movement between electricity-consumption growth and GDP growth, with a coefficient of 0.344. Our estimated coefficient of 0.344 can be placed in the context of the literature examining the relationship between electricity consumption and economic activity. \citet{benchmark_coef}, using panel data for 73 countries, report GDP elasticities with respect to electricity consumption ranging from 0.28 to 0.51 after controlling for common time effects, providing a useful external benchmark for the magnitude of our estimate.    

\begin{table}[H]
\centering
\caption{Normal-times GDP--electricity proxy mapping and GDP-equivalent autarky adjustment}
\label{tab:gdp_main_mapping}
\begin{tabular}{lr}
\toprule
Variable & Estimate \\
\midrule
Normal-times proxy coefficient          & 0.344 \\
HAC standard error                       & 0.069 \\
95\% confidence interval                 & [0.208, 0.480] \\
Observations                             & 43 \\
\midrule
Electricity adjustment (\%)              & -1.824 \\
GDP-equivalent change (\%)               & -0.632 \\
GDP-equivalent loss (EUR bn)             & 7.137 \\
95\% CI for GDP-equivalent loss (EUR bn) & [4.328, 9.940] \\
\bottomrule
\end{tabular}
\begin{minipage}{0.85\textwidth}
\vspace{0.2cm}
\footnotesize
\textit{Notes:} The normal-times coefficient is estimated using
four-quarter log changes in real GDP and temperature-adjusted electricity
consumption, with quarter fixed effects. Observations whose four-quarter
comparison windows overlap the COVID-19 pandemic or the European energy
crisis are excluded. HAC/Newey--West standard errors with four lags and
a small-sample correction are reported. The GDP-equivalent effect is a
proxy-based macroeconomic mapping and should not be interpreted as a
causal GDP or welfare effect. Monetary values are expressed in constant 2024 EUR.
\end{minipage}
\end{table}

As a robustness check, the non-overlapping annual specification yields a coefficient of 0.585 (HC1 standard error: 0.125; $N=10$), compared with 0.344 in the quarterly baseline ($N=43$). Although the annual estimate is larger, both specifications indicate a positive association between electricity consumption and GDP, with overlapping 95\% confidence intervals. Given the limited number of annual observations, this exercise should be interpreted as a sensitivity check rather than definitive evidence of parameter stability. Detailed results are provided in Table~\ref{tab:overlap_robustness}.

Applying this finding to the 1.8\% decrease in electricity consumption previously obtained, we estimate an impact on GDP of about 0.63\%. By 2040, Switzerland’s real GDP is projected to reach approximately €1.13 trillion
in 2024 prices. This estimate is derived from \citet{GDP2040}, who assume
a Swiss GDP of CHF 981.6 billion  in 2017 prices in 2040.\footnote{The CHF 2017 figure is adjusted to 2024 prices using the Swiss CPI (FSO) and converted into euros
using the 2024 average exchange rate of CHF 0.93 per EUR (ECB)} Applying the previously
obtained elasticity on this basis, we estimate an average GDP-equivalent effect of about EUR 7
billion, with an upper bound approaching EUR 10 billion. While such an estimation appears reasonable, many statistical issues and potential biases have to be investigated. While restricting our sample to normal-time periods contributes to stabilizing the coefficient, it also drastically reduces the number of observations. Furthermore, the adaptation period following an autarky policy is likely more similar to an energy crisis era. For those reasons, we estimate Eq. (10) over our full sample, controlling for the impact of COVID-19 and the energy crisis. The results, summarized in Table \ref{tab:gdp_transportability}, address those concerns. First, under this specification, the $\beta$ coefficient is remarkably close to the previous estimate. Second, the coefficient associated with the energy crisis period is not statistically significantly different from the applied estimate, providing no evidence against the stability of this relationship during such disturbed periods. These conclusions support our findings.

\begin{table}[H]
\centering
\caption{Historical transportability of the GDP--electricity proxy mapping}
\label{tab:gdp_transportability}

\small
\setlength{\tabcolsep}{6pt}

\begin{tabular}{lccc}
\toprule
Historical regime & Implied slope & Slope difference & $p$-value \\
\midrule
Normal times              & 0.342 & --     & -- \\
COVID-like period         & 1.721 & 1.379  & $<0.001$ \\
Energy-crisis-like period & 0.299 & -0.043 & 0.836 \\
\bottomrule
\end{tabular}

\vspace{0.2cm}

\begin{minipage}{0.92\textwidth}
\footnotesize
\textit{Notes:} Estimates are obtained from the full-sample
transportability specification with separate intercept and slope
shifts for COVID-like and energy-crisis-like periods.
The slope difference is measured relative to the normal-times slope
estimated within the same interacted specification. The reported
$p$-values correspond to Wald tests of equality between each
exceptional-period slope and the normal-times slope. The preferred
strict normal-times estimate used for the GDP-equivalent calculation
is 0.344.
\end{minipage}

\end{table}

The transportability of our estimate is then assessed by performing Wald tests to assess the statistical similarity of the coefficients across normal, COVID-19, and energy-crisis periods. The null hypothesis is rejected when comparing normal times with the COVID-19 era, whereas it cannot be rejected for the energy-crisis period, suggesting no statistically significant difference from normal times.

\begin{table}[H]
\centering
\caption{Slope-stability tests across exceptional macroeconomic periods}
\label{tab:gdp_main_slope_breaks}

\begin{tabular}{lcc}
\toprule
Null hypothesis & $\chi^2$ statistic & $p$-value \\
\midrule
COVID-like slope = normal-times slope
    & 179.008 & $<0.001$ \\

Energy-crisis-like slope = normal-times slope
    & 0.043 & 0.836 \\

Joint slope stability
    & 257.173 & $<0.001$ \\
\bottomrule
\end{tabular}

\begin{minipage}{0.90\textwidth}
\vspace{0.2cm}
\footnotesize
\textit{Notes:} Wald tests are based on the full-sample
transportability specification with separate intercept and slope
shifts for COVID-like and energy-crisis-like periods.
The null hypothesis in the first two rows is that the corresponding
regime-specific electricity--GDP slope equals the normal-times slope.
The joint test evaluates whether both slope-interaction coefficients
are jointly equal to zero.
\end{minipage}
\end{table}

Finally, as reported in Appendix \ref{app:robust_gdp}, our results are shown to be robust to different specifications (Table \ref{tab:gdp_specification_progression}), to the weather adjustment method (Table \ref{tab:gdp_weather_robustness}), to alternative HAC bandwidths (Table \ref{tab:gdp_hac_robustness}), to alternative definitions of four-quarter growth (Table \ref{tab:gdp_functional_form}) as well as to nonlinearities (Table \ref{tab:gdp_nonlinearity_asymmetry}). Last but not least, as recommended with limited sample size, we want to ensure that our results are not driven or significantly impacted by an outlier. With that aim, we perform a leave-one-out exercise on our main specification. The coefficients from this test are reported in Figure \ref{fig:gdp_leave_one_out}. In the majority of the iterations, the coefficients remain highly stable, lying between 0.32 and 0.36. Only two outliers seem to have a significant impact on the estimate, reaching about 0.4. However, correcting for them would only increase our estimation of the GDP variation. We nevertheless retain these observations in the baseline analysis, both to avoid data-dependent sample selection and to maintain a conservative estimate of the GDP-equivalent effect.

\section{Conclusion} \label{sec:conclusion}

International power market integration is unequivocally one of the most crucial topics for policymakers shaping future energy systems. Although this issue is particularly important for Europe given the large number of countries located within a relatively small geographical area, this strategic issue also concerns other areas and even intercontinental relations, notably with projects aiming at connecting Europe to North Africa. In this regard, this paper aims to shed light on this topic and foster informed and constructive discussion to ultimately follow the most beneficial path for relevant stakeholders. 

To this end, using Switzerland as a realistic and compelling case study, this paper first characterizes the energy landscape the country will face by 2040 based on both existing literature and simulation models. Then, electricity market outcomes (\textit{i.e.}, prices) are derived using a fundamental power market model. We complement the electricity-market welfare analysis with an empirical electricity-demand-based mapping of potential macroeconomic consequences. In doing so, we fill an important gap in the existing literature by investigating extreme scenarios, thus providing estimates for the upper bound of consequences associated with a potential autarky decision.

Our results indicate that a total exclusion from the EU power market could lead to total welfare losses estimated at around EUR 1.5 billion in the reference year alone, with macroeconomic repercussions potentially reaching EUR 7 billion over the same period. The Trade and Autarky scenarios should be interpreted as bounding cases rather than as direct representations of the presence or absence of a Swiss–EU electricity agreement. Although our results strongly support the superiority of market integration over autarky, the analysis relies on several simplifying assumptions, notably concerning retail pricing arrangements, by analyzing day-ahead power prices exclusively. In fact, unlike large customers, many households actually face regulated tariffs that are set in advance for a given year, often with limited within-year variation (e.g., peak/off-peak components). Although a more detailed investigation of retail-pricing arrangements would be valuable, our use of hourly day-ahead prices is unlikely to materially affect our conclusions, since annual retail energy tariffs are typically based on expected procurement costs and therefore tend to track average wholesale price conditions, even if they are not mechanically equal to an ex post quantity-weighted day-ahead average. Moreover, because industrial and services-sector consumers are more exposed to such price variations, relying on hourly wholesale prices provides a closer match to the relevant costs driving economic activity.

Furthermore, various additional consequences that are difficult to estimate have deliberately been set aside. In order to ensure security of supply, any arrangement other than a free-trade agreement would create an additional need for flexible gas-fired power plants. As a result, the autonomy gained by withdrawing from the European power market would entail a greater dependence on volatile and unpredictable world fossil-fuel markets.

Besides the academic interest, our work provides a solid foundation for informed discussion of the future of Swiss energy policy. Indeed, the country recently entered the final phase of long-running negotiations with the European Union regarding several strategic agreements, and the electricity agreement in particular. This forthcoming decision, whether taken by the Swiss Parliament or the population, might be one of the most fundamental and consequential decisions Switzerland will face for decades for its energy system, renewable energy development, and progress toward net-zero $CO_2$ emissions.  

\newpage

\bibliographystyle{aea}
\bibliography{ref}
\newpage

\appendix

\renewcommand{\thetable}{\Alph{section}\arabic{table}}
\renewcommand{\thefigure}{\Alph{section}\arabic{figure}}
\renewcommand{\theequation}{\Alph{section}\arabic{equation}}

\makeatletter
\@addtoreset{table}{section}
\@addtoreset{figure}{section}
\@addtoreset{equation}{section}
\makeatother

\section{Nexus-e assumptions}
\label{app:costs}
\begin{table*}[h]
    \caption{Overnight investment costs for Switzerland's relevant generation and storage technologies in 2040. Lines marked with * reflect the values from \cite{Potencia, Loffler_DIW, EU_ref_scenario_EUREF, Breyer2023_LUW}. Battery costs in EUR/kWh are transformed in EUR/kW assuming energy-to-power ratios of 4 and 2.7 for transmission system (TS) and distribution system (DS), respectively.}
    \centering
    \footnotesize
    \begin{tabular}{ll p{7cm}} \toprule
        Technology   & Inv. cost [EUR/kW] & Source \\ \midrule
        Hydro dam     & 7'231     &~\cite{axpo_power_switcher, Bauer2019}\\
        Hydro pump    & 7'231     &~\cite{axpo_power_switcher, Bauer2019} \\
        Hydro RoR     & 2'890     &~\cite{Potencia, Loffler_DIW, EU_ref_scenario_EUREF, Breyer2023_LUW} \\
        PV - roof     & 1'183     &~\cite{axpo_power_switcher, Bauer2019} \\
        PV - alpine   & 3'091     &~\cite{axpo_power_switcher} \\
        Wind          & 1'427      &~\cite{axpo_power_switcher, Bauer2017} \\ 
        Biomass       & 2'985     &~\cite{axpo_power_switcher}, * \\
        Gas           & 975       &~\cite{axpo_power_switcher} \\
        Gas with CCS  & 1'769     &~\cite{axpo_power_switcher} \\
        Battery TS    & 1'016      &~\cite{Figgener2023, Cole2023} \\ 
        Battery DS    & 1'081      &~\cite{Figgener2023, Cole2023} \\ \bottomrule
    \end{tabular}
    \label{tab:inv_cost}
\end{table*}

\begin{table*}[h]
    \caption{Financing rate by technology based on expert judgment.}
    \centering
    \footnotesize
    \begin{tabular}{ll} \toprule
        Technology   & Financing rate [/] \\ \midrule
        Hydro dam     & 0.0510     \\
        Hydro pump    & 0.0510     \\
        Hydro RoR     & 0.0510     \\
        PV - roof     & 0.0480    \\
        PV - alpine   & 0.0510     \\
        Wind          & 0.0528      \\
        Biomass       & 0.0510     \\
        Gas           & 0.0500       \\
        Gas with CCS  & 0.0500     \\
        Battery TS    & 0.0500      \\ 
        Battery DS    & 0.0500      \\ \bottomrule
    \end{tabular}
    \label{tab:financing_rate}
\end{table*}

\begin{table*}[h]
    \caption{Fixed operating and maintenance (FOM) costs for Switzerland's relevant generation and storage technologies.}
    \centering
    \footnotesize
    \begin{tabular}{ll p{7cm}} \toprule
        Technology    & FOM [EUR/kW] & Source \\ \midrule
        Hydro dam     & 37.51        &~\cite{Potencia, EU_ref_scenario_EUREF, Breyer2023_LUW}\\
        Hydro pump    & 37.51        &~\cite{Potencia, EU_ref_scenario_EUREF, Breyer2023_LUW} \\
        Hydro RoR     & 36.45        &~\cite{Potencia, Loffler_DIW, EU_ref_scenario_EUREF, Breyer2023_LUW} \\
        PV - roof     & 15.00        &~\cite{Lazard2025} \\
        PV - alpine   & 61.82        & BKW professional experience \\
        Wind          & 17.55        &~\cite{Potencia, Loffler_DIW, EU_ref_scenario_EUREF, Breyer2023_LUW, tyndp2022} \\ 
        Biomass       & 0            &~\cite{Bauer2017, Bauer2019} \\
        Gas           & 21.99        &~\cite{Potencia, Loffler_DIW, EU_ref_scenario_EUREF, Breyer2023_LUW} \\
        Gas with CCS  & 46.43        &~\cite{Potencia, EU_ref_scenario_EUREF, Breyer2023_LUW} \\
        Battery TS    & 20.31        &~\cite{Figgener2023, Cole2023} \\ 
        Battery DS    & 21.63        &~\cite{Figgener2023, Cole2023} \\ \bottomrule
    \end{tabular}
    \label{tab:FOM_cost}
\end{table*}
\begin{table*}[h]
    \caption{Non-fuel variable operating and maintenance (VOM) costs for Switzerland's relevant generation and storage technologies. Lines marked with * reflect the values from \cite{Potencia, Loffler_DIW, EU_ref_scenario_EUREF, Breyer2023_LUW}}
    \centering
    \footnotesize
    \begin{tabular}{ll p{7cm}} \toprule
        Technology    & VOM [EUR/MWh] & Source \\  \midrule
        Hydro dam     & 0.87      &~\cite{axpo_power_switcher}, *\\
        Hydro pump    & 0.87      &~\cite{axpo_power_switcher}, * \\
        Hydro RoR     & 2.52      &~\cite{Potencia, Loffler_DIW, EU_ref_scenario_EUREF, Breyer2023_LUW} \\
        PV - roof     & 0.05      &~\cite{Potencia, Loffler_DIW, EU_ref_scenario_EUREF, Breyer2023_LUW} \\
        PV - alpine   & 0         &~\cite{axpo_power_switcher} \\
        Wind          & 0.71      &~\cite{axpo_power_switcher}, * \\ 
        Biomass       & 1         &~\cite{Bauer2017, Bauer2019} \\
        Gas           & 2.74      &~\cite{Potencia, Loffler_DIW, EU_ref_scenario_EUREF, Breyer2023_LUW} \\
        Gas with CCS  & 4.76      &~\cite{Potencia, Loffler_DIW, EU_ref_scenario_EUREF, Breyer2023_LUW} \\
        Nuclear       & 8.80      &~\cite{axpo_power_switcher} \\
        Battery TS    & 0         &~\cite{Cole2023, Kost2024_frauenhofer} \\ 
        Battery DS    & 0         &~\cite{Cole2023, Kost2024_frauenhofer} \\ \bottomrule
    \end{tabular}
    \label{tab:VOM_cost}
\end{table*}

\begin{figure}[H]
    \centering
    \includegraphics[
        width=1\linewidth,
        trim=0 0 0 1.15cm,
        clip
    ]{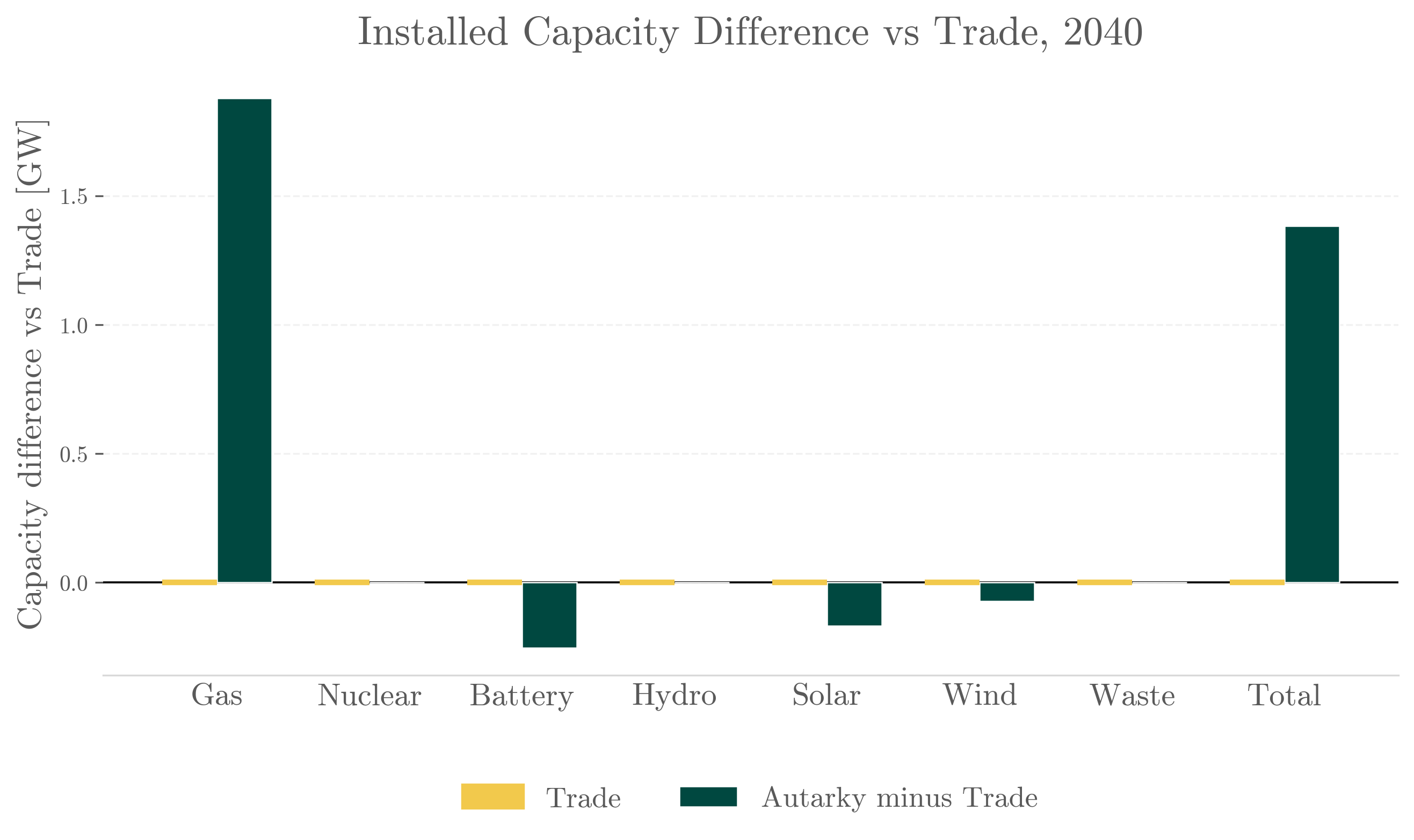}
    \begin{center}
    \caption{Autarky Swiss power generation mix relative to Trade.}
    \label{fig:generation_mix}
    \footnotesize
    \textit{Notes:} This figure shows the difference in generation capacity required in 2040 under autarky relative to Trade to guarantee a comparable security of supply. Most of the difference arises due to the backup gas capacity required in autarky.
    \end{center}
\end{figure}

\begin{figure}[H]
    \centering
    \includegraphics[
        width=1\linewidth,
        trim=0 0 0 1.15cm,
        clip
    ]{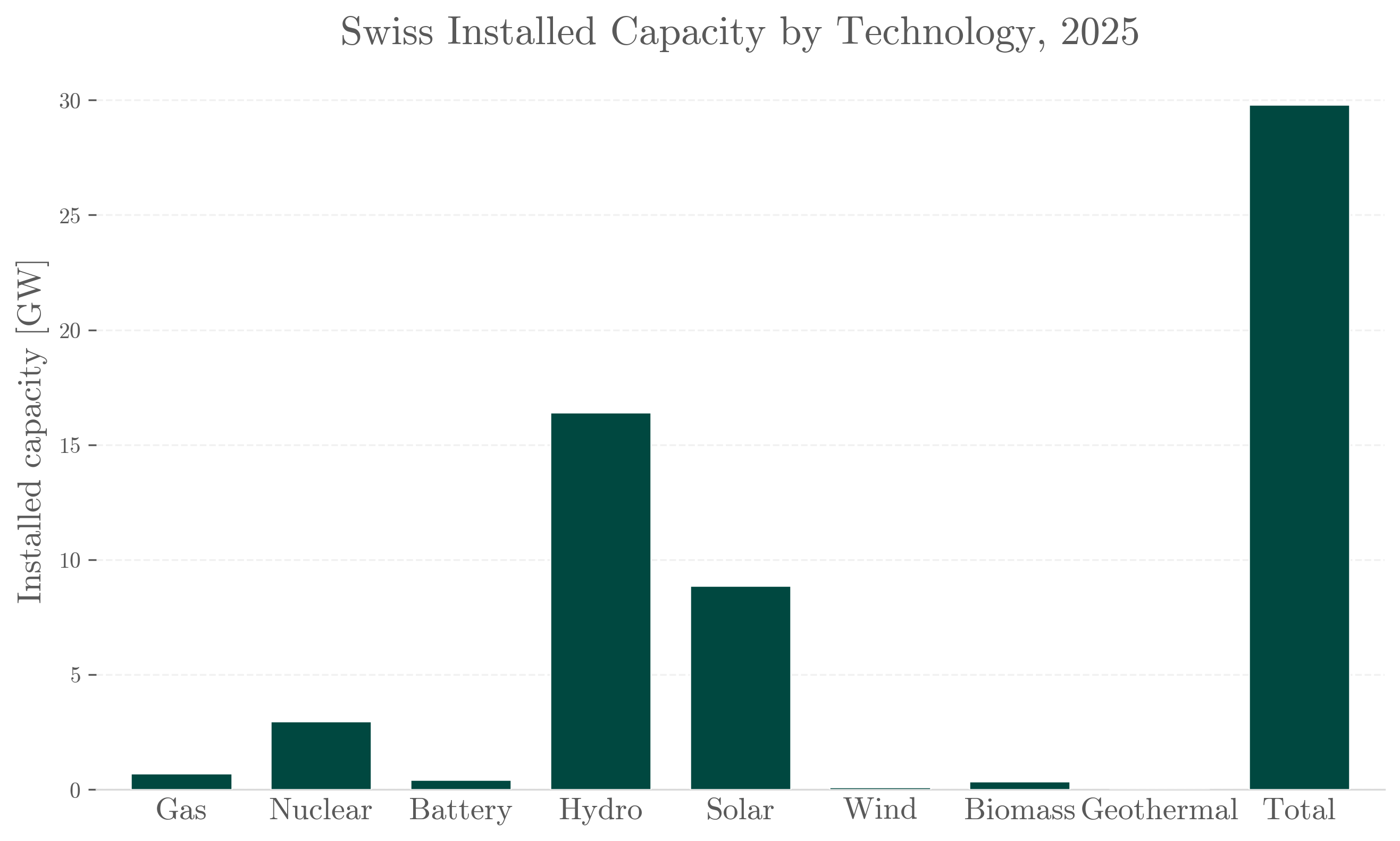}
    \begin{center}
    \caption{Swiss power generation mix 2024.}
    \label{fig:generation_mix_start}
    \footnotesize
    \textit{Notes:} This figure shows the generation capacity in 2024, which serves as the starting point for the Nexus-e model.  
    \end{center}
\end{figure}

\newpage
\section{Comparison with modeling results from other Swiss-focused studies}
\label{app:other_results}

\begin{figure}[H]
    \centering
    \includegraphics[width=1\linewidth]{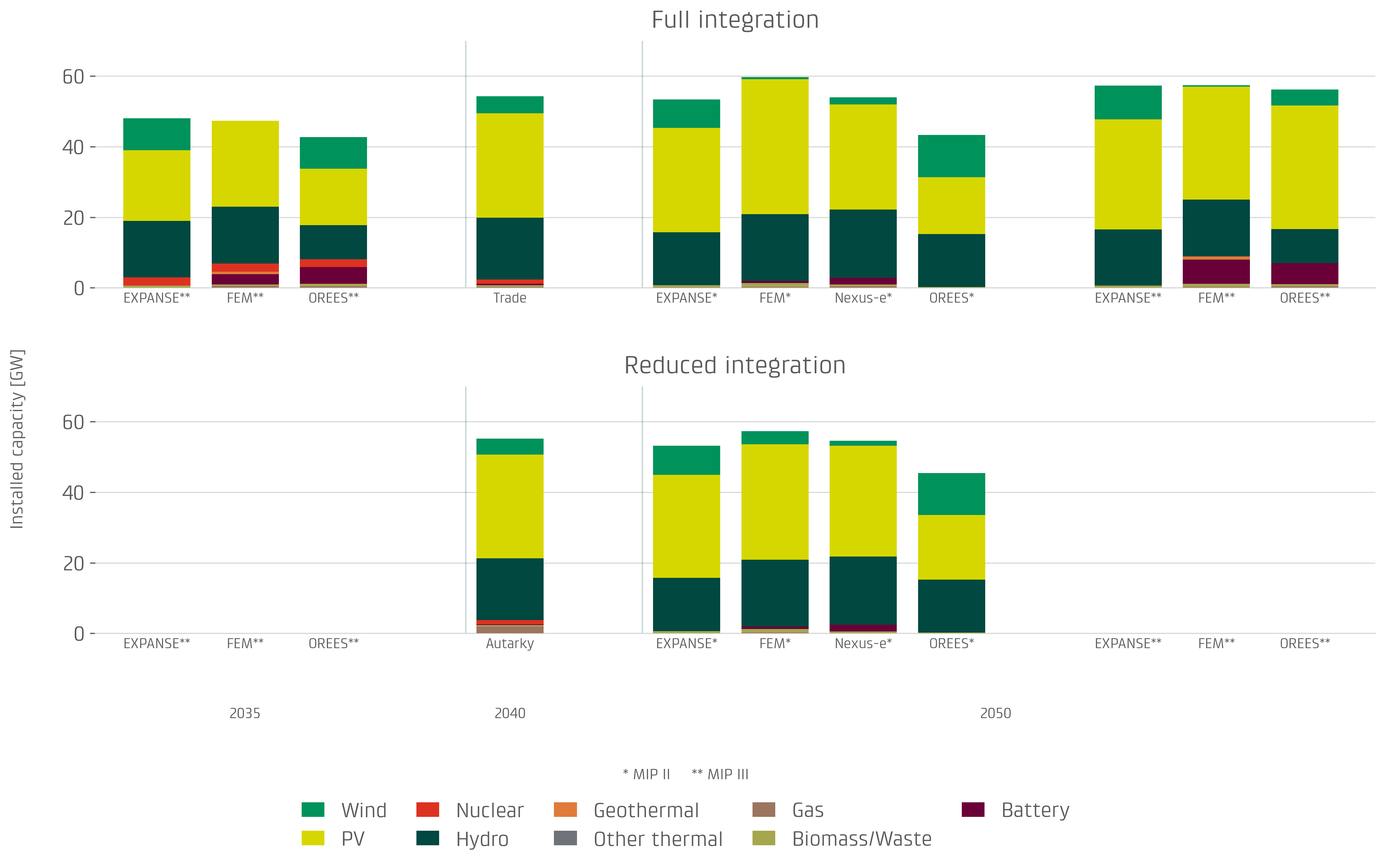}
    \caption{Generation mix for selected scenarios from the second (*) and third (**) SWEET EDGE model inter-comparison studies (\cite{van_liedekerke_policy-relevance_2024, kerscher2026}), compared to the Trade and Autarky scenarios.}
    \label{fig:generation_mix_EDGE}
\end{figure}

In the SWEET EDGE model inter-comparison studies, the CentIv model is applied alongside other models to evaluate a range of future scenarios covering different national energy policies and technology developments (\cite{Heinisch2023, van_liedekerke_policy-relevance_2024, kerscher2026}). 
These results provide useful context for the present study.
Figure~\ref{fig:generation_mix_EDGE} shows results from the second and third model inter-comparisons (denoted * and **, respectively), which examine scenarios similar to those considered here. 
The second inter-comparison assesses the effects of a renewable generation target combined with reduced market integration in 2050, the latter modeled as a reduction of net transfer capacity (NTC) to 30\% of its full value. 
The third inter-comparison instead considers the years 2035 and 2045 under a renewable generation target with no demand-side flexibility.
Figure~\ref{fig:generation_mix_EDGE} shows that, although generation mixes across the full integration scenarios vary, they are comparable.
Additionally, it shows that reduced integration requires different costly and sub-optimal investments in gas and/or wind generation. 

\begin{figure}[h]
    \centering
    \includegraphics[width=0.75\linewidth]{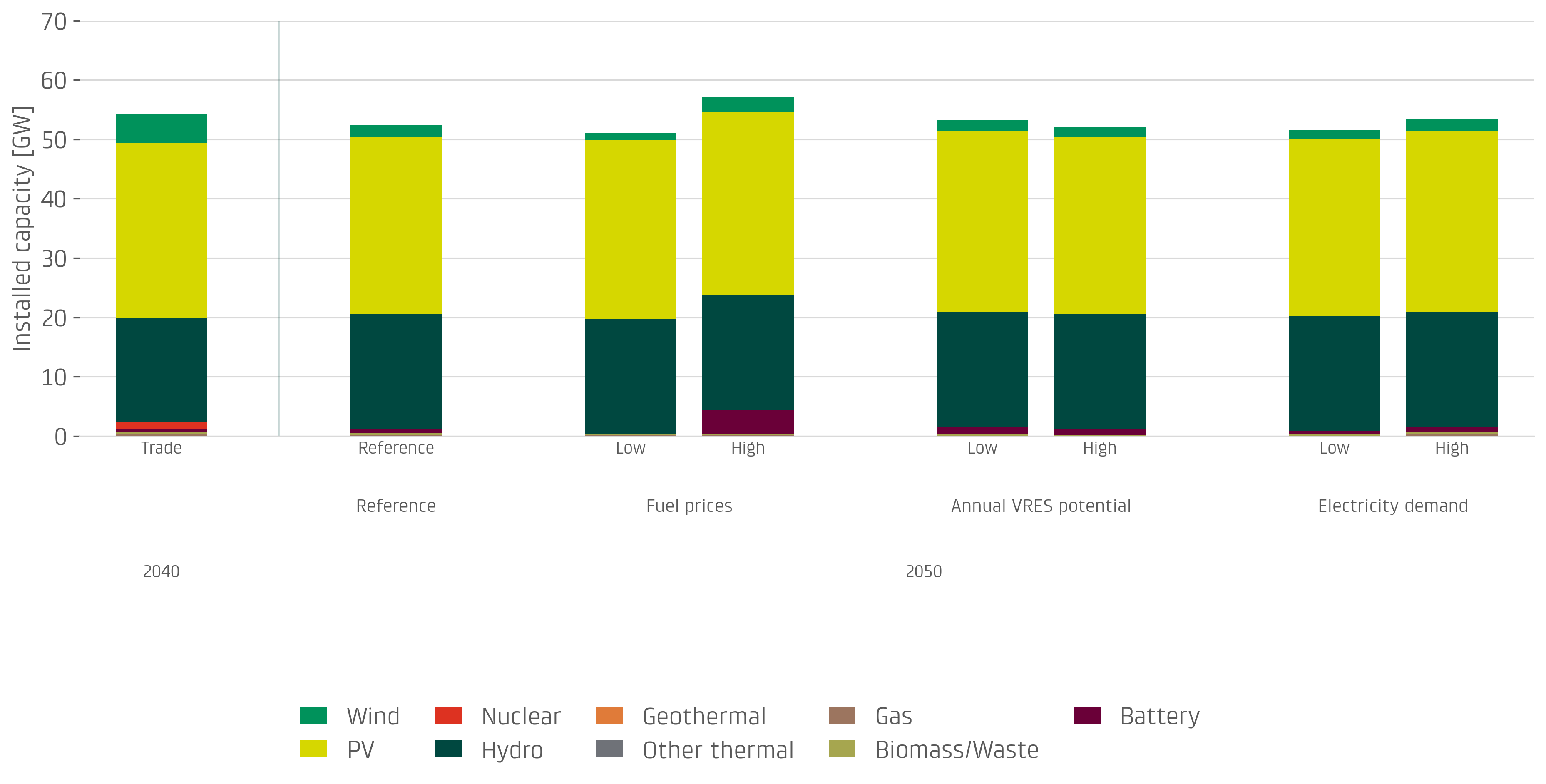}
    \caption{Generation mix for selected scenarios from~\cite{VanLiedekerke2026_ags}, compared to the Trade scenario.}
    \label{fig:generation_mix_AGS}
\end{figure}

The CentIv model is also applied in several other studies. 
In~\cite{VanLiedekerke2026_ags}, it is used to explore the potential deployment of a specific geothermal technology in Switzerland. 
Within this context, a range of scenarios is examined, considering low and high values for a series of input parameters, including fuel prices, renewable generation output, and electricity demand, in order to assess their impact on the resulting generation mix. 
The corresponding generation mixes are shown in Figure~\ref{fig:generation_mix_AGS}.

\newpage
\section{Welfare analysis}
\label{app:robust}

\begin{figure}[H]
    \centering
    \includegraphics[
        width=0.9\linewidth,
        trim=0 0 0 1.15cm,
        clip
    ]{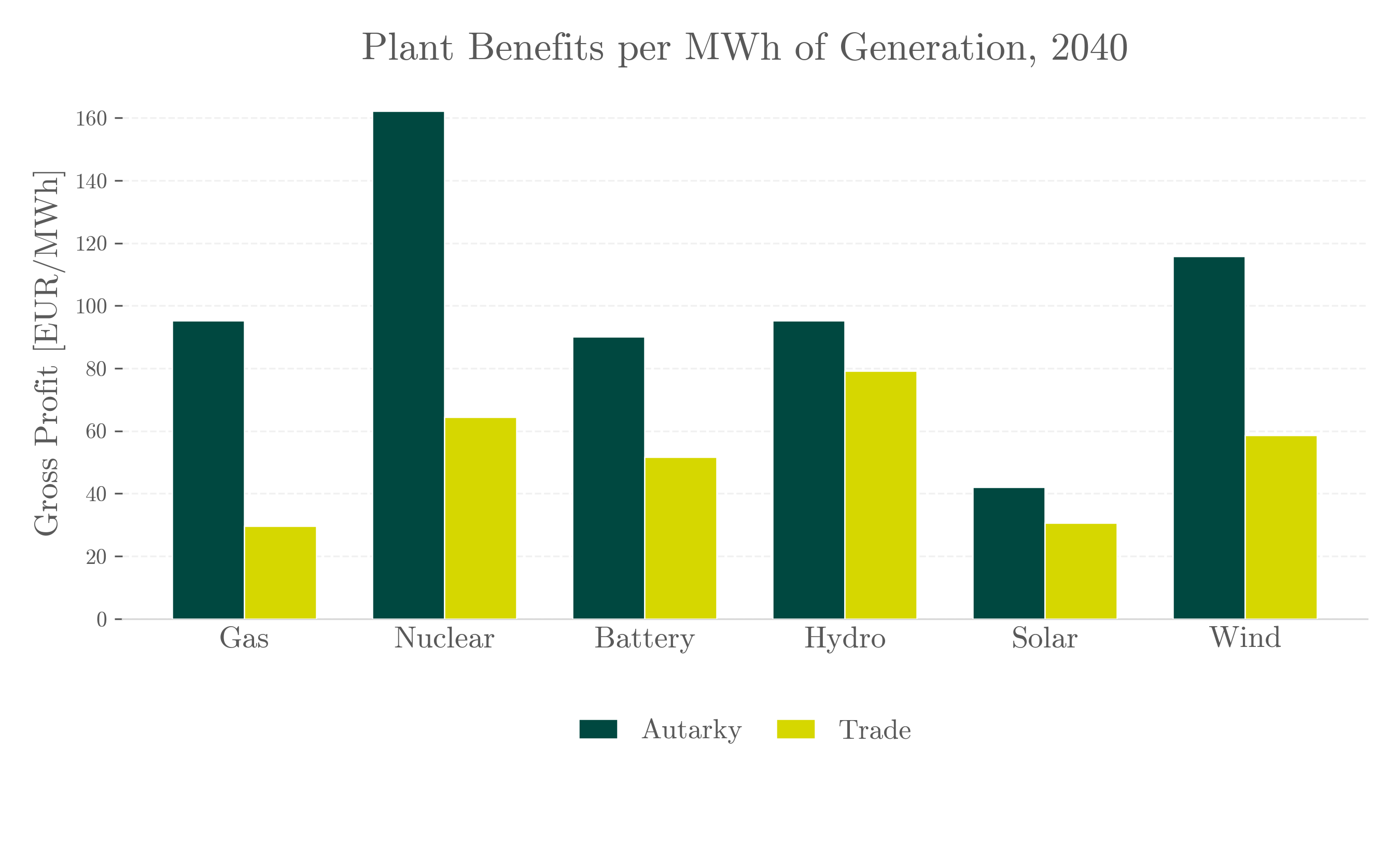}
    \begin{center}
    \caption{Technology-wise 2040 Gross Profit by MWh.}
    \label{fig:ProducerBenefitsMWh}
    \footnotesize
    \textit{Notes:} This figure shows Gross Profit by MWh generated for the different technologies used under the two scenarios. Monetary values are expressed in constant 2024 EUR.
    \end{center}
\end{figure}

In a context of rising geopolitical tensions and volatile political positions with regard to climate policy and climate change, fuel and $CO_2$ prices are among the most uncertain assumptions of this paper. Consequently, this section investigates the consequences of such variations on welfare. Starting from the central case used throughout the analysis, we allow for a 33\% variation in those parameters as summarized in Table \ref{tab:Fuels_prices}.

As shown in Table \ref{tab:economic-benefit-analysis_sensitivity}, prices have a significant impact on overall welfare gains from trade. However, large variations as set in the sensitivity analysis do not alter the qualitative conclusion of this study, showing the significant robustness of our results.

\begin{table}[htbp] 
\centering 
\begin{threeparttable} 
\caption{Fuel Prices} 
\label{tab:Fuels_prices} 
\begin{tabular}{llrrrr} 
\toprule Scenarios & $CO_2$ & $H_2$ & Gas & Coal & Nuclear\\ 
\midrule 
High Prices  & 195.5 & 145.5 & 47 & 20.9 & 9 \\
Reference Prices & 147 & 109.4 & 35.3 & 15.7 & 6.8 \\
Low Prices  & 97 & 72.2 & 23.3 & 10.4 & 4.5 \\ 
\bottomrule 
\end{tabular} 
\begin{tablenotes}
\footnotesize 
\item 
\textit{Note:} The prices are expressed in 2024 EUR per tonne for $CO_2$ and in 2024 EUR per MWh for fuels. \end{tablenotes}
\end{threeparttable} 
\end{table}

\begin{table}[htbp] 
\centering 
\begin{threeparttable} 
\caption{Economic Benefit Analysis (Million 2024 EUR)} 
\label{tab:economic-benefit-analysis_sensitivity} 
\begin{tabular}{llrrrr} 
\toprule Scenarios & $\Delta$ CS & $\Delta$ PS & $\Delta$ INV & $\Delta$ GRID & Total Welfare \\ 
\midrule 
High Fuel Prices  & 4457 &  -2685   &    95   &     23   &      1890 \\
Reference Fuel Prices & 3189  & -1826   &    95    &    23    &     1481 \\
Low Fuel Prices  & 2062  & -1115   &    95    &    23    &     1065 \\ 
\bottomrule 
\end{tabular} 
\begin{tablenotes}
\footnotesize 
\item 
\textit{Note:} Positive values indicate a welfare gain under Trade relative to Autarky. \end{tablenotes}
\end{threeparttable} 
\end{table}

\section{GDP impact mapping}
\label{app:robust_gdp}

\begin{table}[H]
\centering
\caption{Robustness of the GDP--electricity relationship across specifications}
\label{tab:gdp_specification_progression}

\small
\setlength{\tabcolsep}{4pt}

\begin{tabular}{lcccccc}
\toprule
Specification
& $\hat{\beta}$
& HAC SE
& \multicolumn{2}{c}{95\% CI}
& Adj. $R^2$
& $N$ \\
\cmidrule(lr){4-5}
& & & Lower & Upper & & \\
\midrule
Full sample + quarter FE
& 0.642 & 0.271 & 0.110 & 1.174 & 0.299 & 63 \\

+ Crisis intercepts
& 0.684 & 0.267 & 0.161 & 1.206 & 0.300 & 63 \\

+ Separate slope breaks
& 0.342 & 0.070 & 0.205 & 0.479 & 0.612 & 63 \\

+ Linear trend
& 0.339 & 0.086 & 0.171 & 0.507 & 0.605 & 63 \\

Strict normal-times sample
& \textbf{0.344} & \textbf{0.069}
& \textbf{0.208} & \textbf{0.480}
& 0.186 & 43 \\
\bottomrule
\end{tabular}

\begin{minipage}{0.95\textwidth}
\vspace{0.15cm}
\footnotesize
\textit{Notes:} The dependent variable is the four-quarter log change
in real GDP. Electricity consumption is measured by the four-quarter
log change in temperature-adjusted total electricity demand.
HAC/Newey--West standard errors use four lags.
The strict normal-times specification excludes observations whose
four-quarter comparison windows overlap the COVID-19 pandemic or the
European energy crisis.
\end{minipage}

\end{table}

\begin{table}[H]
\centering
\caption{Robustness to alternative weather adjustments}
\label{tab:gdp_weather_robustness}

\small
\setlength{\tabcolsep}{5pt}

\begin{tabular}{lccccc}
\toprule
Weather adjustment
& $\hat{\beta}$
& HAC SE
& \multicolumn{2}{c}{95\% CI}
& $N$ \\
\cmidrule(lr){4-5}
& & & Lower & Upper & \\
\midrule
Baseline
    & 0.344 & 0.069 & 0.208 & 0.480 & 43 \\
Month fixed effects
    & 0.332 & 0.077 & 0.182 & 0.482 & 43 \\
Month FE + linear trend
    & 0.344 & 0.061 & 0.224 & 0.463 & 43 \\
\bottomrule
\end{tabular}

\begin{minipage}{0.90\textwidth}
\vspace{0.15cm}
\footnotesize
\textit{Notes:} Each specification uses total electricity consumption
(24/7) and the strict normal-times sample. The baseline weather
adjustment controls for average temperature, squared temperature, and
high- and low-temperature deviations. The alternative specifications
add month fixed effects and, respectively, a linear time trend to the
weather-adjustment equation. HAC/Newey--West standard errors use four
lags.
\end{minipage}

\end{table}

\begin{table}[H]
\centering
\caption{Robustness to alternative HAC bandwidths}
\label{tab:gdp_hac_robustness}

\small
\setlength{\tabcolsep}{6pt}

\begin{tabular}{ccccc}
\toprule
HAC lags
& $\hat{\beta}$
& SE
& \multicolumn{2}{c}{95\% CI} \\
\cmidrule(lr){4-5}
& & & Lower & Upper \\
\midrule
3
& 0.344 & 0.074 & 0.199 & 0.489 \\

4 (baseline)
& 0.344 & 0.069 & 0.208 & 0.480 \\

8
& 0.344 & 0.060 & 0.226 & 0.462 \\
\bottomrule
\end{tabular}

\begin{minipage}{0.88\textwidth}
\vspace{0.15cm}
\footnotesize

\textit{Notes:} The table reports Newey--West HAC standard errors under alternative bandwidths for the preferred strict normal-times specification, with four lags in the baseline. All specifications include quarter fixed effects. The decline in standard errors at higher bandwidths reflects negative weighted autocovariance contributions from the regression scores. Since crisis-window exclusions introduce gaps in the sample, HAC lags refer to consecutive retained observations rather than consecutive calendar quarters.

\end{minipage}

\end{table}

\begin{table}[H]
\centering
\caption{Robustness to alternative definitions of four-quarter growth}
\label{tab:gdp_functional_form}

\small
\setlength{\tabcolsep}{6pt}

\begin{tabular}{lcccc}
\toprule
Growth measure
& $\hat{\beta}$
& HAC SE
& \multicolumn{2}{c}{95\% CI} \\
\cmidrule(lr){4-5}
& & & Lower & Upper \\
\midrule
Log changes (baseline)
& 0.344 & 0.069 & 0.208 & 0.480 \\

Exact percentage changes
& 0.350 & 0.070 & 0.213 & 0.487 \\
\bottomrule
\end{tabular}

\begin{minipage}{0.90\textwidth}
\vspace{0.15cm}
\footnotesize
\textit{Notes:} Both specifications are estimated on the strict
normal-times sample and include quarter fixed effects.
The first specification uses four-quarter log changes in real GDP
and temperature-adjusted electricity consumption. The second uses
the corresponding exact percentage changes. HAC/Newey--West
standard errors use four lags.
\end{minipage}

\end{table}

\begin{table}[H]
\centering
\caption{Tests for nonlinearity and asymmetric responses}
\label{tab:gdp_nonlinearity_asymmetry}

\small
\begin{tabular}{lccc}
\toprule
Specification
& Coefficient(s)
& Null hypothesis
& $p$-value \\
\midrule
Linear baseline
& $\beta=0.344$
& --
& -- \\

Quadratic
& $\beta_1=0.338,\;\beta_2=0.747$
& $\beta_2=0$
& 0.831 \\

Positive/negative slopes
& $\beta_+=0.415,\;\beta_-=0.247$
& $\beta_+=\beta_-$
& 0.512 \\
\bottomrule
\end{tabular}

\begin{minipage}{0.92\textwidth}
\vspace{0.15cm}
\footnotesize
\textit{Notes:} All specifications use the strict normal-times sample
and include quarter fixed effects. The quadratic specification tests
for curvature in the GDP--electricity relationship. The piecewise
specification allows the slope to differ between increases and
decreases in electricity consumption. Reported $p$-values correspond
to Wald tests of the stated null hypotheses. HAC/Newey--West standard
errors use four lags.
\end{minipage}

\end{table}

\begin{figure}[H]
    \centering
    \includegraphics[width=0.95\textwidth]{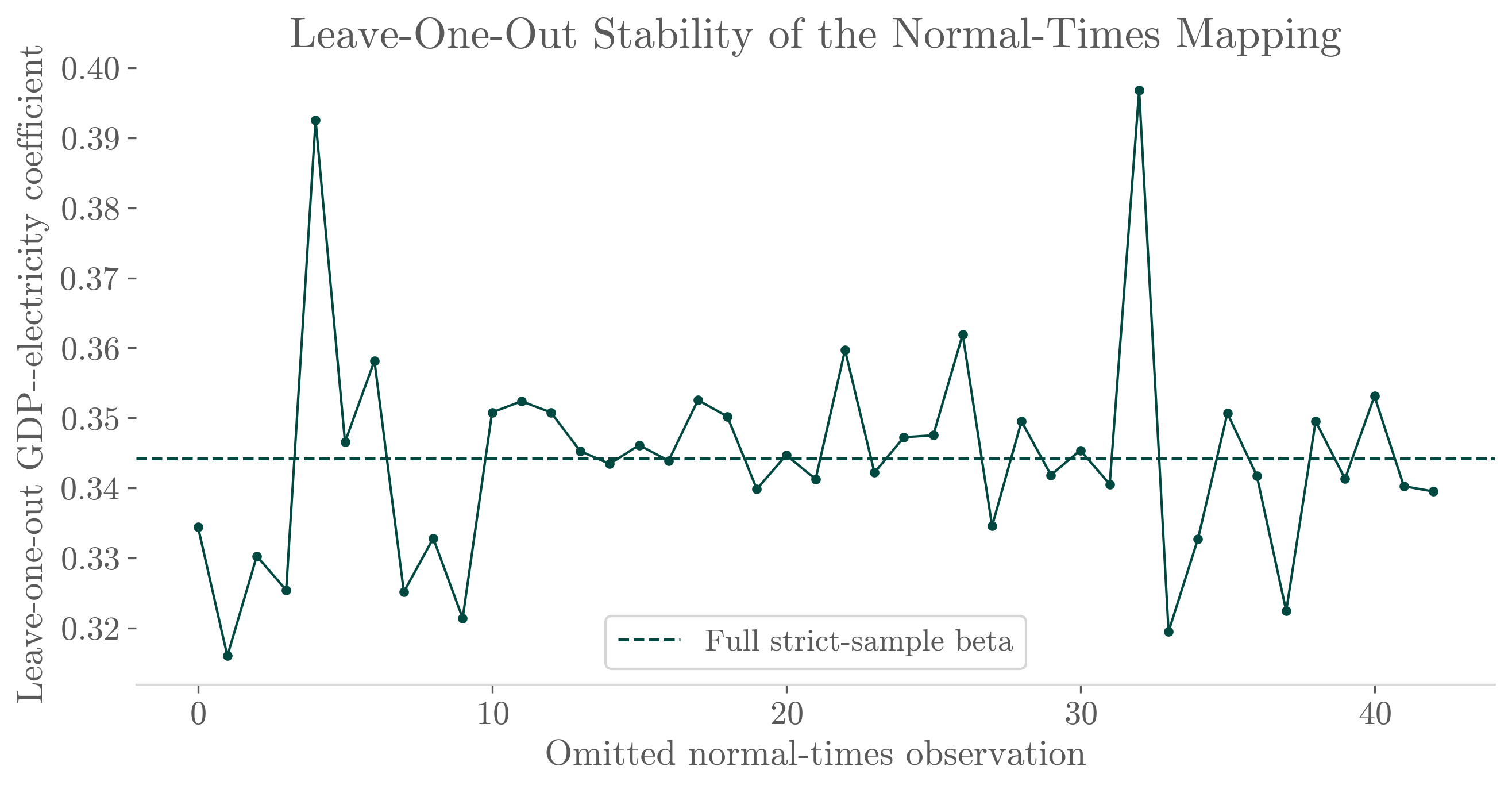}
    \caption{Leave-one-out stability of the normal-times GDP--electricity mapping.}
    \label{fig:gdp_leave_one_out}
\end{figure}

\begin{table}[htbp]
\centering
\caption{Robustness to overlapping four-quarter observations}
\label{tab:overlap_robustness}
\begin{tabular}{lcc}
\hline\hline
 & Quarterly baseline & Annual Q4-to-Q4 \\
\hline
Coefficient $\hat{\beta}$ & 0.3442 & 0.5846 \\
Standard error & 0.0693 & 0.1251 \\
$t$-statistic & 4.97 & 4.67 \\
95\% confidence interval
    & [0.208, 0.480] & [0.296, 0.873] \\
Observations & 43 & 10 \\
Standard error estimator & HAC(4) & HC1 \\
Overlapping observations & Yes & No \\
\hline\hline
\end{tabular}

\vspace{0.2cm}
\begin{minipage}{0.95\textwidth}
\footnotesize
\textit{Notes:}
The dependent variable is the four-quarter log change
in real GDP, and the explanatory variable is the
corresponding change in temperature-adjusted
electricity consumption. The quarterly specification
includes quarter fixed effects and uses Newey--West
HAC standard errors with four lags.
The annual specification retains one Q4-to-Q4
observation per year over 2010--2019 and reports
HC1 heteroskedasticity-robust standard errors.
The annual sample eliminates mechanical overlap
between consecutive growth observations but remains
small, limiting statistical inference.
\end{minipage}
\end{table}
\end{document}